# Exploring the Mechanical Behaviors of 2D Materials in Electrochemical Energy Storage Systems: Present Insights and Future Prospects

**Dibakar Datta**

Department of Mechanical and Industrial Engineering
New Jersey Institute of Technology, Newark, NJ 07103, USA
Email: dibakar.datta@njit.edu

## Abstract

2D materials (2DM) and their heterostructures (2D + $n$D, $n$ = 0,1,2,3) hold significant promise for applications in Electrochemical Energy Storage Systems (EESS), such as batteries. 2DM can serve as van der Waals (vdW) slick interface between conventional active materials (e.g., Silicon) and current collectors, modifying interfacial adhesion and preventing stress-induced fractures. Additionally, 2DM can replace traditional polymer binders (e.g., MXenes). This arrangement also underscores the critical role of interfacial mechanics between 2DM and active materials. The intercalation-deintercalation (charging-discharging) process profoundly influences the interfacial behavior and various mechanical properties of 2DM when employed as binders or current collectors. Furthermore, 2DM can be designed to function as an electrode itself. For instance, a porous graphene network has been reported to possesses approximately five times the capacity of a traditional graphite anode. Consequently, gaining a comprehensive understanding of the mechanical properties of 2DM in EESS is paramount. However, modeling 2DM in EESS poses significant challenges due to the intricate coupling of mechanics and electrochemistry. For instance, defective graphene tends to favor adatom adsorption (e.g., Li+) during charging. In cases of strong adsorption, adatoms may not readily detach from electrodes during discharging. As a result, in such scenarios, adsorption-desorption (charge-discharge) processes govern the mechanical properties of 2DM when used as binders and current collectors. Regrettably, most existing studies on the mechanical properties of 2DM in EESS have failed to adequately address these critical issues. This perspective paper aims to provide a comprehensive overview of recent progress in the chemo-mechanics of 2DM's mechanical properties. It will elucidate the challenges associated with modeling 2DM in EESS and outline potential future research directions. A wide spectrum of multiscale modeling approaches, including atomistic/molecular simulations, continuum modeling, and machine learning, are discussed. This perspective paper intends to stimulate further research in the mechanics of 2DM in EESS and offer valuable guidelines for experimentalists seeking to optimize the design of 2DM-based EESS for real-life applications.

## 1. INTRODUCTION

In contemporary society, there exists a pressing demand for highly efficient and cost-effective solutions to promote the widespread adoption of rechargeable batteries across various applications, including portable electronics, grid storage, renewable energy storage, and electric vehicles[1]. Batteries have become an integral part of our daily lives[2]. Most rechargeable battery products rely on Lithium-Ion Batteries (LIBs), which were honored with the 2019 Nobel Prize in Chemistry[2]. However, the primary drawback of lithium-ion technology is its cost. Lithium, a finite resource on our planet, has experienced a reported price surge of approximately 738% since January 2021[2]. As an alternative to lithium, earth-abundant and cost-effective metals like aluminum (Al), calcium (Ca), magnesium (Mg), and Zinc (Zn) have been actively researched for battery systems[3-7]. These metals are multivalent, meaning they carry two or more ionic charges. Consequently, one ion insertion results in the delivery of two or more electrons per ion during battery operation, offering the promise of enhanced performance and cost efficiency. In summary, both mono- and multivalent ions are under extensive investigation for sustainable energy storage.

However, the existing intercalation hosts pose several challenges. For instance, the conventional graphite anode used in LIBs provides a mere gravimetric capacity of 372 mAh/g. In contrast, a silicon anode can achieve a high capacity of 4000 mAh/g. Nevertheless, a significant obstacle in utilizing silicon anodes for battery operations is the substantial volume expansion of 300% during lithiation[8]. In the realm of beyond lithium batteries, higher charge density multivalent ions exhibit robust interactions with ions in the intercalation host, resulting in a relatively high migration barrier and slower diffusion kinetics. The insertion of multivalent ions contributes to the mechanical degradation of the hosts, leading to stress-induced cracking and a loss of material integrity[2]. Consequently, both mono- and multivalent ion-based batteries confront substantial challenges in identifying suitable hosts for energy storage.

Over the past few decades, the field of two-dimensional (2D) materials has experienced remarkable growth, offering significant promise in the realm of energy storage[9-11]. This promise stems from their substantial specific surface areas, which provide electrochemically active sites for ion storage, as well as open channels that facilitate rapid ion transport[12]. 2D materials serve as a bridge between one-dimensional (1D) and three-dimensional (3D) bulk materials, introducing novel fundamentals challenges associated with low-dimensional materials and creating a wealth of new applications. Their intrinsic mechanical flexibility and the prevalent van der Waals (vdW) bonding in 2D materials make them exceptionally suitable for seamless integration with existing active materials, ultimately enhancing battery performance. Due to their diverse properties, 2D materials have the potential for applications in various facets of batteries, including anodes, cathodes, conductive additives, electrode-electrolyte interfaces, separators, and electrolytes. Researchers have conducted numerous studies to explore the utilize of 2D materials and their heterostructures for energy storage[13].

Extensive research efforts have delved into various mechanical properties of 2D materials, including fracture[14-16], friction[17, 18], and interlayer shear[19], among others[20-23]. Elastic properties such as young's modulus and stiffness have been determined through both experimental and theoretical approaches for various 2D materials[24, 25]. Different modes of fracture have undergone through investigation, and computational and experimental studies have scrutinized friction, shear, and other mechanical aspects of 2D materials and their heterostructures[20, 23]. Numerous perspective and review papers have comprehensively discussed these critical facets of 2D materials[14, 26-28].

However, despite the extensive study of the mechanical properties of 2D materials and their heterostructures, several challenges arise when applying them in Electrochemical Energy Storage Systems (EESS), such as batteries[13]. In EESS, the behavior of 2D materials is influenced by several factors, including intercalation ions (e.g., Li, Na, Ca, Mg, K), charge/discharge rates, and various types of electrolytes (organic, aqueous), among others. For instance, most, if not all, studies on graphene fracture fail to consider these factors. The impact of intercalation, electrolyte choice, and charge/discharge rates on fracture has yet to be explored comprehensively in both computational and experimental studies. A similar knowledge gap exists in other crucial mechanical studies, such as interlayer friction, shear, stability, and wrinkling. Furthermore, various properties of 2D/3D interfaces (e.g., graphene/silicon) have undergone extensive examination, including interface adhesion and interfacial charge transfer[29-31].

Nonetheless, there is a notable absence of extensive studies on interfaces that consider ion intercalation, charge/discharge rates, and electrolyte conditions, among other pertinent factors. Perhaps most importantly, the study of the mechanical properties of 2D materials in EESS present several challenges. Conducting Density Functional Theory (DFT) calculations[9, 10] on these systems proves complicated due to the substantial computational costs involved. While Molecular Dynamics (MD)[8] studies can be valuable, the essential interatomic potential for MD simulations of 2D materials in EESS remains unavailable for many systems. Although Machine Learning (ML)[32, 33] modeling has seen extensive use in studying the mechanical properties of 2D materials[34], implementing ML models for 2D materials in EESS systems is a challenging endeavor. The primary obstacle is the lack of training data; all existing databases (e.g., Materials Project[35], C2DB[36], OQMD[37]) consist solely of individual materials data, with no consideration for homo-/hetero-structures, intercalated systems, charge-discharge rates, electrolytes, etc. The generation of training data for the development of ML models would be a time-consuming task. Consequently, there exists a substantial opportunity for the mechanics community to explore these intricate and challenging problems.

In this prospective, we aim to address the challenging issues and enormous opportunities of studying the mechanical properties of 2D materials in energy storage systems. The structure of the paper is organized as follows: **Section 2** discusses the existing problems associated with battery architecture and different failure modes that are commonly observed. **Section 3** highlights how 2D materials and their heterostructures can be implemented to overcome some of the common failure patterns observed in batteries. This section focuses on the potential benefits and advantages of using 2D materials for improving battery performance and reliability. **Section 4** discusses the potential problems with mechanics of 2D materials in energy storage systems. Several emerging topics are discussed such as fracture, interface, homo- and heterostructures, phase transition, interlayer friction and layer-dependent fracture, nano-confined fluid-induced mechanics, wrinkling, solid electrolyte interface (SEI) formation. **Section 5** provides a summary of the existing computational challenges and future directions. **Finally**, paper concludes summarizing the novel key take-home messages. We believe that this prospective review will stimulate further theoretical and experimental work in mechanical properties of 2D materials in electrochemical energy storage systems.

## 2. PROBLEMS WITH THE EXISITING ENERGY STORAGE SYSTEMS

### 2.1 Traditional Battery Architecture

Figure 1 depicts the conventional architecture of a LIB, and it's worth mentioning that batteries relying on other ions share a similar structure[13]. While this review primarily centers on LIBs as a model system, the insights and discussion can be readily applied to other ion-based batteries as well. LIBs consist of two key components: a positive electrode and a negative electrode, both separated by an electrolyte. The electrolyte can take the form of an organic solution, such as ethylene carbonate (EC), or an aqueous solution, for example, Lithium bis(trifluoromethanesulfonyl)mide (LiTFSI). In the discharging process, lithium ions (cations) are stored in the negative electrode, typically composed of graphite, migrate through the electrolyte, and insert themselves into the positive electrode material, such as $LiCoO_2$. Simultaneously, external electrons flow towards the positive electrode. During the charging process, this movement of ions and electrons is reversed.

To prevent electrical short circuits between the electrodes while facilitating the unhindered diffusion of lithium ions, an ionically transparent and electrically insulating separator is positioned between the positive and negative electrodes. Historically, polymers have been employed as binders in batteries, and materials like copper, nickel, and others are commonly used as current collectors. It's essential to recognize that while this review primarily examines the LIB architecture mentioned above, the fundamental principles discussed can be extrapolated to batteries based on different ions.

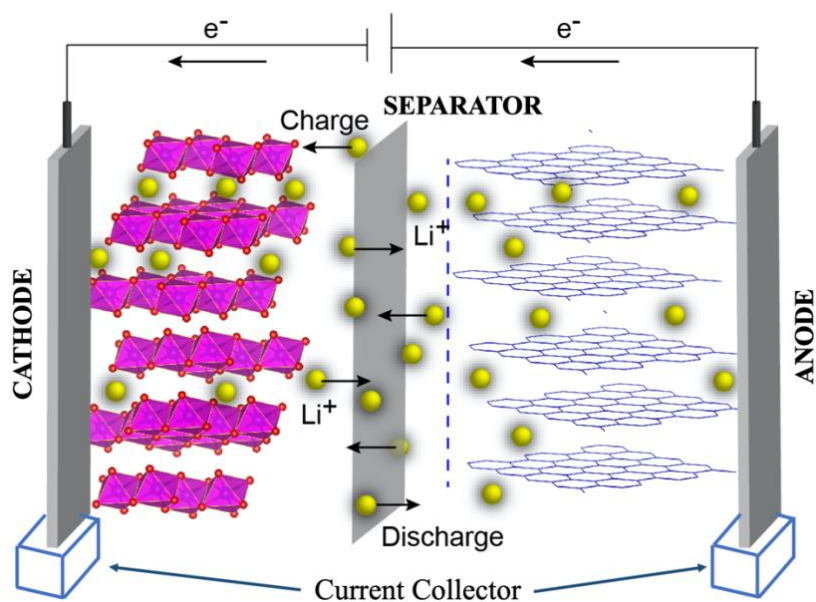

**Figure 1: Traditional battery configuration**. The electrodes (cathode and anode) are placed over the current collector. Reprinted with permission from ref[38]. Copyright 2018 @Royal Society of Chemistry.

### 2.2 Typical Battery Failure Modes

Batteries exhibit various failure mechanisms that warrant dedicated investigation. Among these, two prominent chemo-mechanical failure modes are particularly noteworthy, as illustrated in Figure 2.

***(i) Interface Failure Leading to Electrical Isolation of Active Particles:***

The active particles[39, 40], such as Silicon (Si), contact metal current collectors (Figure 2b1), typically composed of materials like nickel (Ni). These current collectors ensure a uniform distribution of electrons[41]. However, during the intercalation process, the active particles undergo substantial volume expansion[42]. For example, Si can expand by up to 300% upon lithiation[8, 43]. The metal current collectors, acting as non-flexible surfaces, cannot accommodate this volume expansion and contraction of the active particles. This results in the generation of excessive interfacial stress (Figure 2b1), leading to the fracture of active particles and ultimately causing battery failure[44].

Another critical interface within the battery system exists between the polymer binder and active particles[40] (Figure 2b2). The polymer binder network plays a crucial role in holding active particles together and providing a pathway for electron transport throughout the electrode[45]. However, the volume expansion and contraction of active particles during charging and discharging generate excessive interfacial stress and subsequent fracture, leading to the electrical isolation of active particles[40] (Figure 2b2). When the active particles become electrically isolated, they can no longer contribute to the battery's overall electrochemical reactions, resulting in a decline in battery performance.

Addressing these issues related to interfacial stress and the fracture of active particles is paramount for improving the performance and reliability of batteries. Strategies aimed at mitigating these challenges are actively pursued to enhance the overall functionality and lifespan of battery systems.

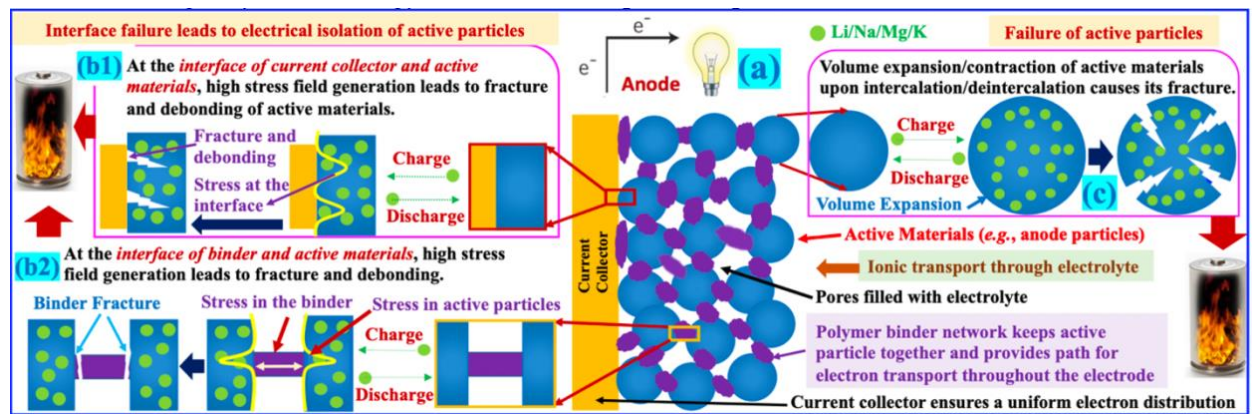

**Figure 2: Current failure modes in battery: [a]** Schematic of anode, **[b]** Different interface failure modes, **[c]** Failure of active particles.

***(ii) Failure of Active Particles:***

In addition to interface failure, the fracture of active particles is a significant factor contributing to battery failure[46-50]. For instance, the volume expansion and contraction experienced by Si anode particles during lithiation and delithiation cycles can lead to their fracture[51], ultimately resulting in battery failure[52]. Finding effective strategies to address these practical challenges is of utmost importance.

## 3. 2D MATERIALS FOR ENERGY STORAGE

### 3.1 Overview of 2D Materials

Figure 3 illustrates various 2D materials and their heterostructures (2D + $n$D, where $n$ = 0, 1, 2, 3), each with a wide array of applications spanning multiple fields. The inception of the 2D materials field can be traced back to the isolation of monolayer graphene flakes from bulk graphite through mechanical exfoliation[53]. Since then, numerous 2D materials have been discovered, including transition metal dichalcogenides (TMDs) such as $MoS_2$[54], hexagonal boron nitride (h-BN)[55], black phosphorous (BP) (or phosphorene)[56], and MXenes (*e.g.*, $Ti_3C_2$)[57]. The 2D materials family boasts a diverse set of physical properties. For instance, graphene exhibits outstanding conductivity, while materials like $MoS_2$ demonstrate semiconducting behavior, and h-BN exhibits insulating properties[26, 58]. By the year 2020, over a thousand 2D materials had been identified[59], opening new avenues for groundbreaking physics and applications across various domains, including energy storage.

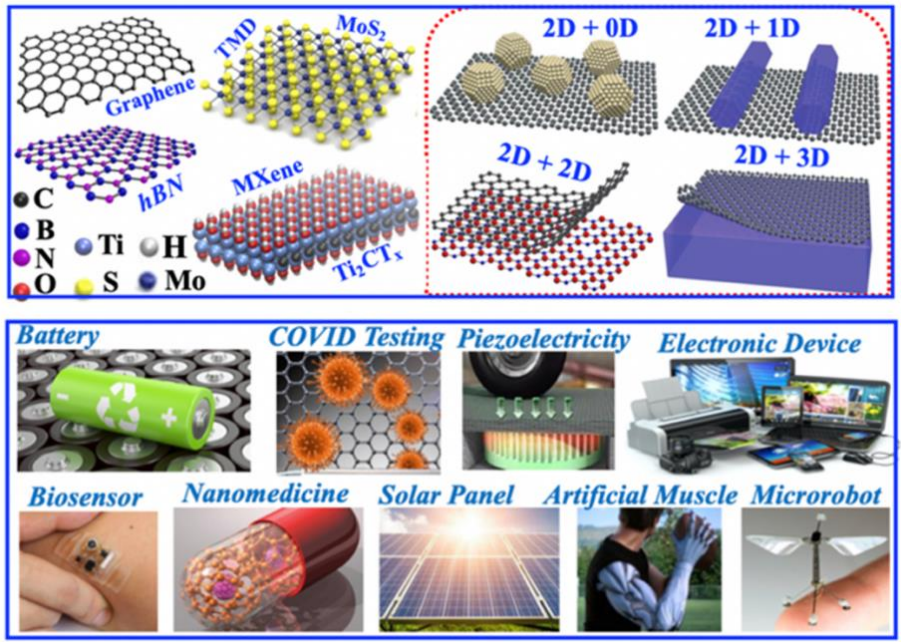

**Figure 3: Different 2D materials** – graphene, hBN, transition metal dichalcogenides (TMDs), MXene. 2D+$n$D heterostructures, n = 0,1,2,3, and various applications of 2D materials.

### 3.2 Overview of How 2D Materials Benefit Battery Applications

Figure 4 presents a comparative analysis between electrodes based on 3D materials and those based on 2D materials concerning their performance in batteries. In the context of a 3D electrode architecture, battery capacity typically experiences a decline over time, and the electrode's resistance tends to increase. This deterioration can result in reduced overall battery performance and a shorter operational lifespan. Conversely, within a 2D electrode scenario, when appropriate design strategies are applied, it becomes feasible to preserve the battery's capacity over time. This implies that 2D materials hold significant potential for a wide range of applications in battery design. The comparison illustrated in Figure 4 underscores the advantages afforded by 2D materials, particularly in their ability to sustain battery capacity and performance throughout the battery's lifespan. These findings strongly suggest that the incorporation of 2D materials into battery electrode designs can lead to substantial improvements in battery performance and longevity[13, 60-62].

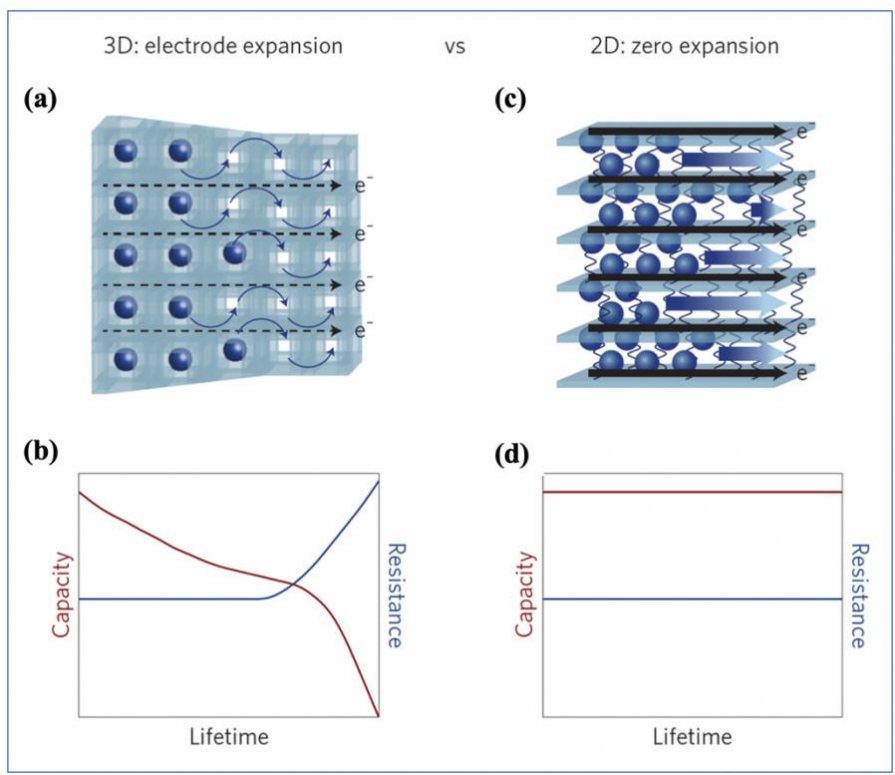

**Figure 4: Overcoming limitations of current batteries by using 2D materials. [a]** Schematic of a 3D intercalation electrode, and **[b]** increase in resistance and decrease in capacity because of mechanical and/or structural degradation of electrode materials. **[c]** A zero-expansion 2D electrode with improved kinetics of electrochemical processes due to easy transport of electrons and ions, and **[d]** stable electrochemical performance and extended lifetime due to the close-to-zero volume change of pillared 2D electrodes. Reprinted with permission from ref[63]. Copyright @ Nature.

### 3.2.1 Prevention of Interface Failure

**2D Materials as Binder:**

One potential solution to address the challenges associated with the polymer binder in battery electrode is the substitution of traditional binders with 2D MXenes (Figure 5a1)[64]. We hypothesize that 2D materials, like MXenes, can serve as a 'slippery' surface, effectively reducing interfacial stress. By replacing the conventional polymer binder with 2D MXenes, it becomes possible to minimize friction and stress at the interface between the active particles and the binder. The unique properties of MXenes, including their high mechanical strength and flexibility, position them as promising candidates for binder materials. These 2D materials offer improved adhesion and mechanical stability, lowering the risk of stress-related failures[44].

**2D Materials as Current Collector:**

To tackle interface failure between the current collector and active particles (Figure 2b1), we propose two potential options:

*(i) Addition of a 2D Materials as a 'Coating' on the Current Collector:* Figure 5a2 illustrates this approach[44]. By applying a thin layer of graphene or another suitable 2D materials onto the current collector's surface, it can function as a lubricating layer, reducing interfacial stress and friction between the current collector and the active particles[31]. This method enhances mechanical compatibility and alleviates stress concentration at the interface.

*(ii) Complete Replacement of the Current Collector with 2D Materials:* This alternative is represented in Figure 5a3[65]. In this scenario, the traditional current collector material, such as copper or nickel, would be entirely substituted with a 2D material. For example, MXenes[30] is renowned for their excellent conductivity and mechanical properties can serve as both the current collector and structural support for the active materials. This replacement eliminates the interface between the current collector and active particles, potentially reducing interfacial stress and improving overall electrode performance. Both options aim to mitigate the interface failure between the current collector and the active particles by leveraging the unique properties of 2D materials. Whether by coating the current collector or completely replacing it with a 2D material, these approaches can help alleviate stress concentration and enhance the mechanical stability of the electrode interface, ultimately improving battery performance and reliability.

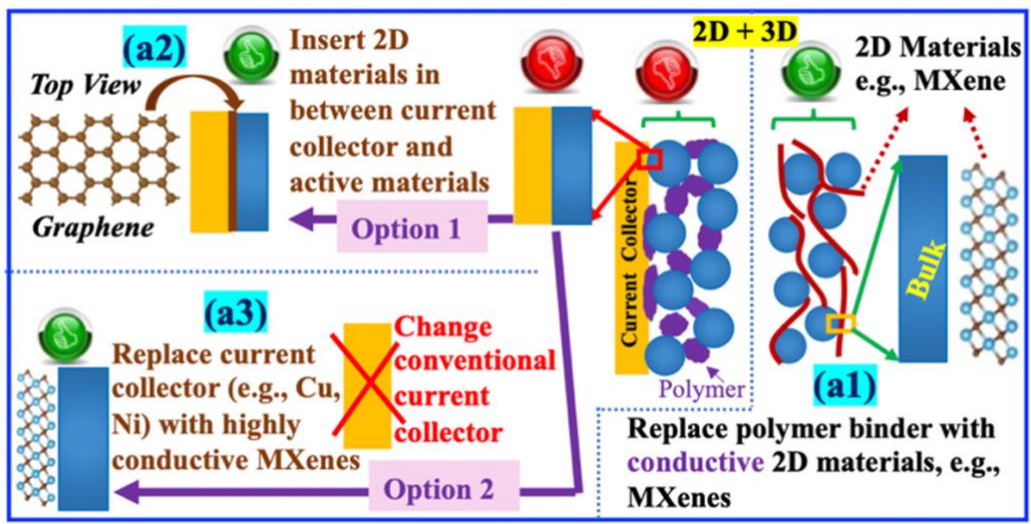

**Figure 5: 2D materials as current collector and binder.**

### 3.2.2 Prevention of Active Materials Failure

To address the challenges of active materials failure (Figure 2c), we propose two distinct approaches:

***(i) Replace 3D Active Materials with 2D Materials and Their heterostructures:*** As illustrated in Figure 6b1, this approach involves substituting 3D active materials (e.g., Si) with suitable 2D materials[11] and their heterostructures (e.g., MXenes + $MoS_2$, Graphene + $MoS_2$). By making this replacement, it becomes possible to potentially overcome the limitations and failure modes associated with 3D materials. 2D materials[9, 10] offer unique properties that can enhance the stability, mechanical flexibility, and electrochemical performance of battery electrodes.

***(ii) Integrated 3D Active Materials with 2D Materials:*** In this approach, as shown in Figure 6b2, we do not entirely replace the 3D active materials. Instead, we combine them with 2D materials[66]. By integrating 3D active materials with suitable 2D materials, we can harness the strengths of both material types. The 2D materials can serve as a protective layer, coating, or interface modifier, thereby enhancing the stability, mechanical properties, and electrochemical performance of the 3D active materials.

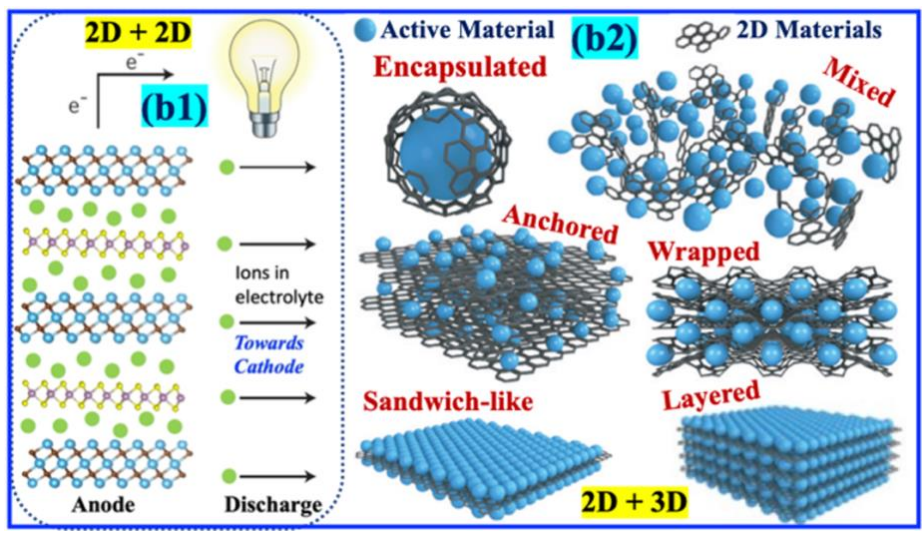

**Figure 6: 2D materials for anode, current collector, binder: [a]** 2D materials as current collector and binder, **[b]** 2D layers and 2D composite as anode materials.

In all these proposed approaches, the role of interfacial mechanics is crucial in determining electrochemical performance metrics such as energy and power density, volumetric capacity, and more. The mechanical integrity of these interfaces (e.g., 2D+2D, 2D+3D) fundamentally influences the long-term performance of energy storage systems[67, 68].

### 3.3 2D Materials as Electrode

Figure 7 demonstrates the utility of 2D graphene in energy storage applications[11], particularly as an electrode material. The mechanism involves defect-induced storage of metallic lithium within a porous structure network (PGN), resulting in a high-performance electrode material. Figure 7a illustrates the PGN electrode, which is central to the experimental setup. The experimental voltage-capacity plot highlights the remarkable performance of PGN (Figure 7b). It achieves a capacity of up to 1000 mAh/g, nearly three times higher than that of conventional graphite (372 mAh/g). These experimental findings have been corroborated through theoretical research. DFT (Density Functional Theory) calculations on graphene are depicted in Figure 7c, including pristine graphene (Figure 7c1), divacancy defect (Figure 7c2), and stone-Wales defect (Figure 7c3). Pristine graphene is not conductive to adatom adsorption. However, defects in graphene facilitate the adsorption of adatoms, enhancing its capacity for lithium storage. Figure 7d illustrates that the capacity of a defective graphene-based electrode can be three to four times higher than that of conventional graphite. These results underscore the potential of 2D materials, such as graphene, in significantly improving the performance and capacity of electrodes in energy storage applications, offering a promising avenue for advancements in battery technology.

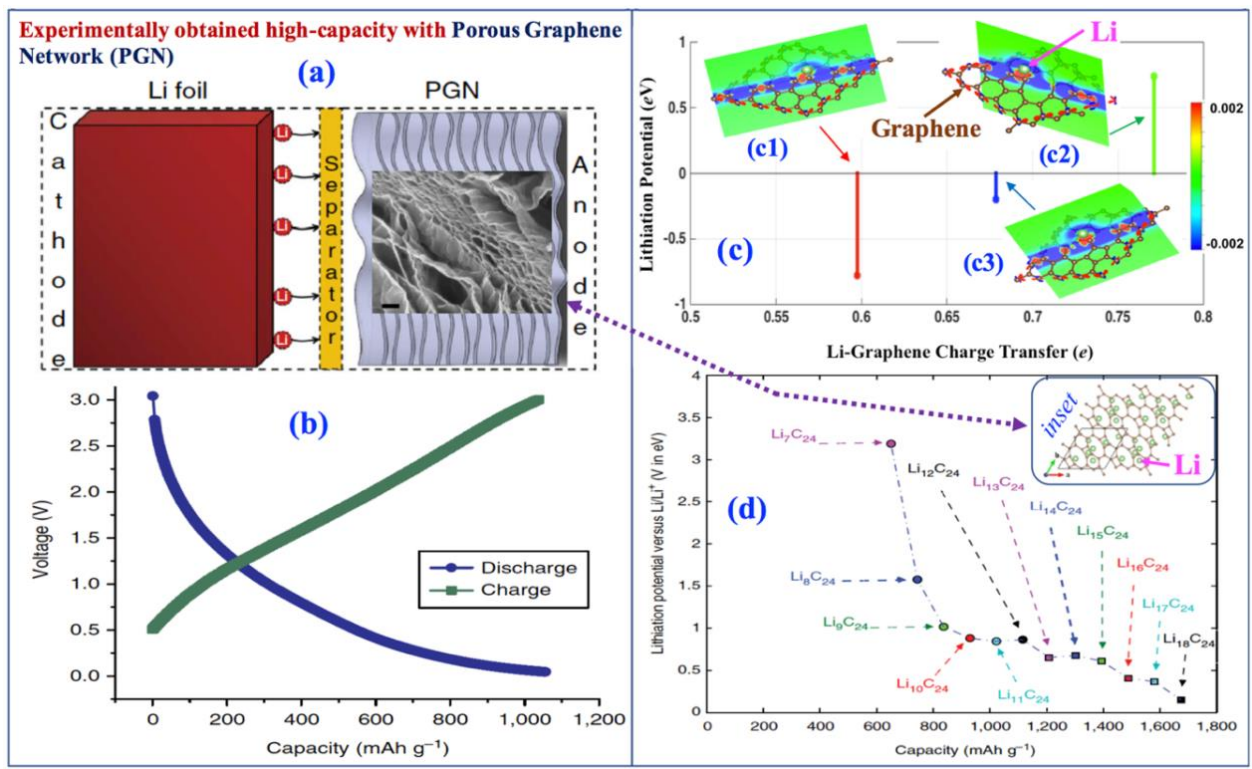

**Figure 7: Defective graphene as high-capacity anode materials.** [**a-b**] Experimental work showing the porous graphene network (PGN) for high-capacity anode. [**c**] DFT study of Li adsorption on graphene – (c1) pristine, (c2) stone-wale, (c3) divacancy. [**d**] DFT study of capacity-potential curve for defective graphene. Reprinted with permission from ref[11]. Copyright 2014 @Nature.

## 3.4 2D Materials Integrated with Active Materials

*Graphene-Enhanced Alloy-Type Anode Materials:*

In conventional silicon-based lithium batteries, when silicon undergoes lithiation and transforms into a $Li_{4.4}Si$ alloy, it experiences significant volume expansion (320%). This expansion can lead to particle pulverization[69], which poses challenges for electrode stability and overall battery performance[46]. To address these issues, many researchers have incorporated 2D materials like graphene[70-75]. Luo et al.[76] developed denser crumpled graphene balls containing silicon particles (Figure 8a). In Figures 8b and 8c, it is evident that the Si@crumpled graphene anode material exhibited significantly improved Initial Coulombic Efficiency (ICE) and capacity compared to bare silicon nanoparticles during the first 20 cycles. This improvement can be attributed to the protective effect of graphene on the anode surface. Without the stabilizing influence of graphene, the Solid Electrolyte Interphase (SEI) formed on silicon nanoparticles tends to repeatedly fracture and rebuild with each cycle. This process consumes a substantial amount of lithium ions and electrolyte and results in the formation of a thick and poorly conductive SEI layer. In contrast, the Si@crumpled graphene anode outperformed the pristine silicon anode in terms of cycling reversibility due to the surface stabilization facilitated by graphene. These findings demonstrate the potential of 2D materials, such as graphene, to enhance the stability and performance of anode materials in lithium batteries, ultimately contributing to improved battery longevity and efficiency.

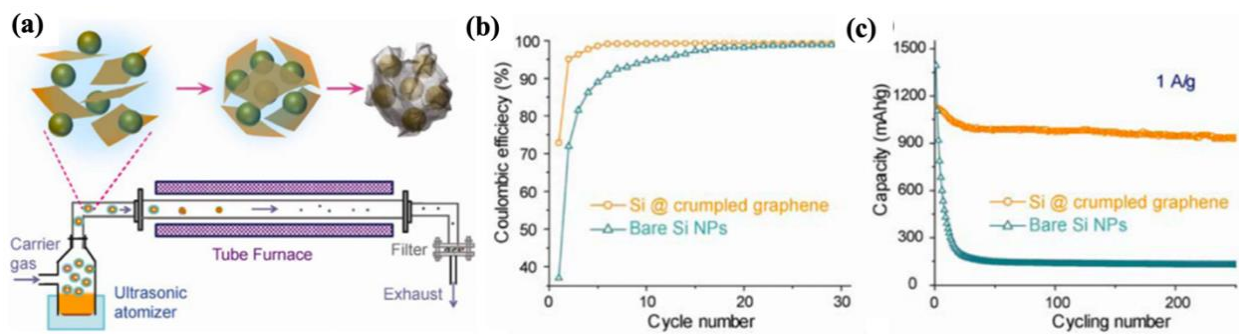

**Figure 8:** **[a]** Schematic drawing illustrating the synthesis of the crumpled-graphene-wrapped Si nanoparticles. **[b]** Coulombic efficiency of the Si@crumpled graphene compared to the bare Si nanoparticles. **[c]** Charge/discharge cycling data of the Si@crumpled graphene compared to the bare Si nanoparticles. Reprinted with permission from ref[76]. Copyright 2012 @American Chemical Society.

Li et al.[74] employed silicon microparticles as the anode materials, a choice known for its disadvantages, including structural pulverization and uncontrolled SEI growth after lithiation (Figure 9a)[46]. This mechanism represents the classic failure of micron-sized silicon anode materials, resulting in the formation of inactive silicon clusters after cycling. To address these challenges, the authors developed a mechanically flexible graphene cage designed to accommodate the silicon microparticles while preserving void space (Figure 9b). While the lithiated silicon microparticles still experienced fracture, the fragmented particles were confined within the conformal graphene cage. Consequently, the electrical contact of the active silicon was maintained, and the SEI only grew on the surface of the graphene. The defects present on the graphene cage served as pathways for lithium-ion transport, which were subsequently sealed by the formation of a thin SEI layer after cycling. This innovative approach allowed for improved stability and performance of silicon microparticles as anode materials, representing a significant step toward enhancing the overall efficiency and longevity of lithium batteries[77].

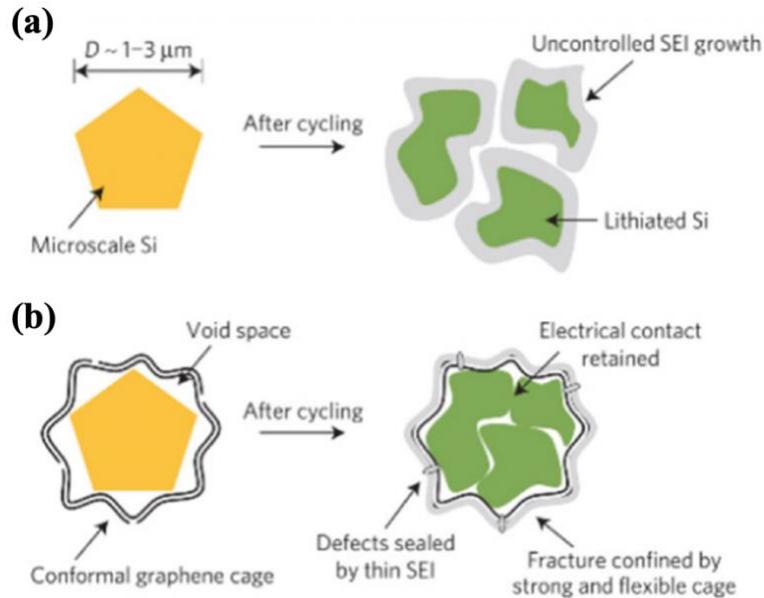

**Figure 9: [a]** The failure mechanism of Si microparticles in the anode during repeated battery cycling. **[b]** Enhanced structural stability and SEI control of Si microparticles with a mechanically flexible graphene cage. Reprinted with permission from ref[74]. Copyright 2016 @ Nature.

Mo et al. developed a 3D N-doped graphene foam to encapsulate germanium particles (Figure 10)[78]. This design aimed to create a yolk-shell structure with internal void space. The inclusion of extra void space serves to buffer the volume expansion that occurs when germanium alloys with lithium, thereby controlling the size of secondary particles. However, it's important to note that if the density of anode particles is too low, achieving a high volumetric energy density may become challenging. Additionally, the gap between the internal active material (germanium) and the graphene can lead to poor conductivity, which can in turn affect the rate capability of the electrode material. Balancing these factors is crucial when designing anode materials for lithium batteries to optimize both volumetric energy density and rate performance.

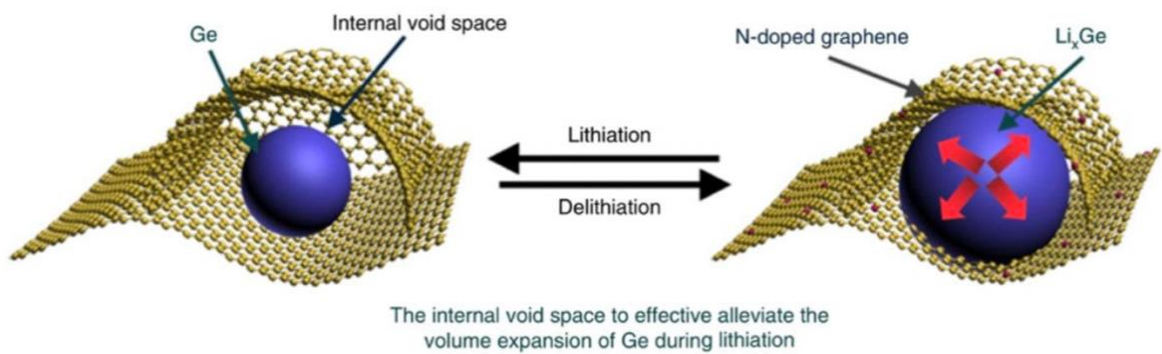

**Figure 10:** Schematic drawing of the lithiation/delithiation process of the graphene-Ge yolk-shell composite anode material. Reprinted with permission from ref[78]. Copyright 2017 @ Nature.

*Graphene-Enhanced Lithium Metal and Lithiated Anode Materials*

Wang et al.[79] developed a graphene cage embedded with gold particles to serve as a container for storing lithium metal (Figure 11a-d). This innovative design involves lithium deposition and stripping within the graphene cage matrix. Since gold is a lithiophilic seeding material, lithium metal prefers precipitating around the gold particles. As a result, lithium metal continues to deposit and fill the internal space of the graphene cage. During the delithiation process, the lithium metal is stripped away, while the graphene cage architecture remains intact. The graphene cage design exhibits exceptional structural stability, leading to an impressive cycle life of over 300 cycles in lithium metal batteries when an appropriate high-concentration LiFSI electrolyte formulation is used simultaneously. Even when employed alone without the LiFSI electrolyte, the graphene-enhanced lithium metal anode can still extend the cycle life to at least three times longer.

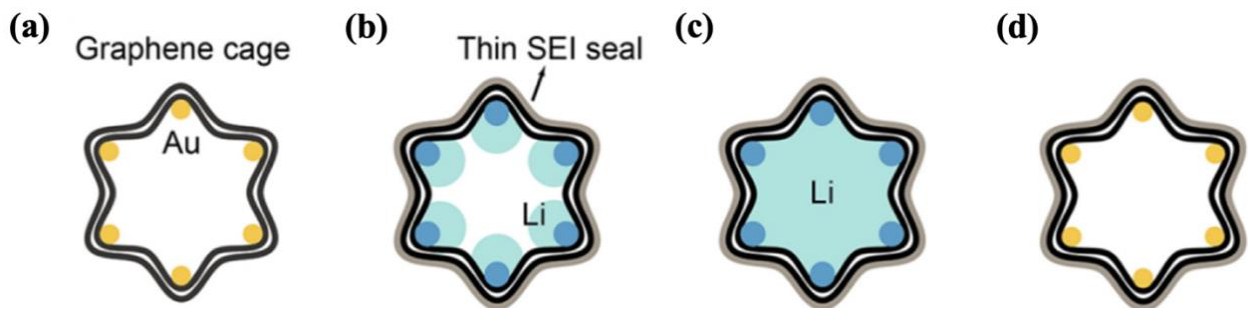

**Figure 11:** Comparison between the graphene cage and copper foil during Li deposition and stripping. **[a]** Pristine graphene cage. **[b,c]** The graphene cage after various amounts of Li deposition. **[d]** The graphene cage after Li stripping. Reprinted with permission from ref[79]. Copyright 2019 @American Chemical Society.

Graphene can protect lithiated anode materials. Figure 12a illustrates a graphene-protected lithiated silicon anode that effectively rejects gas penetration[80]. Gases present in the air, such as water, oxygen, and carbon dioxide, can react with lithium metal and lithiated compounds, leading to the formation of LiOH, $Li_2O$, and $Li_2CO_3$. These reactions can reduce the reversible anode capacity when the materials are assembled in cells. In contrast to a bare $Li_xSi$ anode, the graphene encapsulated $Li_xSi$ exhibited remarkable stability in terms of capacity retention. After two weeks of storage in a dry environment, it retained more than 90% of its initial capacity.

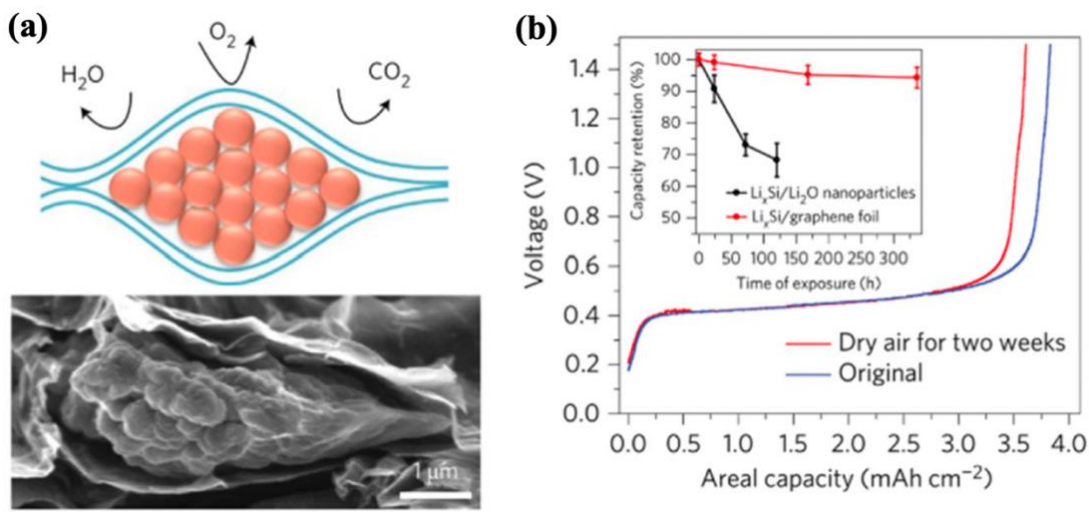

**Figure 12: [a]** Top: Schematic drawing of graphene sheets to avoid gas penetration. Bottom: Cross-sectional SEM image of $Li_xSi$ nanoparticles encapsulated by the graphene sheets. **[b]** The capacity retention of the $Li_xSi$/graphene composite anode in the dry room for two weeks. Reprinted with permission from ref[80]. Copyright 2017 @ Nature.

### 3.5 2D Materials as van der Waals (vdW) Slippery Interface Over Current Collectors

Current collectors play a crucial role in influencing battery performance, and aluminum and copper foils are commonly chosen as current collectors for cathodes and anodes due to their relatively low cost and high electrical conductivity. However, there are several issues associated with these commonly used current collectors: *(i) Corrosion of Aluminum (Al) Collectors:* Al current collectors are prone to corrosion, which can lead to an increase in internal resistance, significant self-discharge, and even potential micro-short circuits caused by Al fragments during extensive cycling[81, 82], particularly at higher potentials. *(ii) Poor Adhesion:* Adhesion between current collectors and electrode materials can be inappropriate, leading to performance issues. *(iii) High Mass Ratio:* The mass ratio of current collectors in the battery is relatively high, typically in the range of 9-10 wt%, which can impact the overall battery performance[83].

In silicon-based batteries, the large volume expansion of silicon during cycling can result in stress build-up at the interface between the silicon film and the current collector. This stress can lead to delamination of the silicon from the current collector surface. Traditionally, adhesion promoters, such as chromium interlayers, have been used to enhance the interface between silicon and the current collector to address this issue. However, recent research by Basu et al.[44] suggests that a more effective approach is to engineer a van der Waals (vdW) "slippery" interface between the silicon film and the current collector. This can be achieved by coating the current collector surface with graphene sheets. In this type of interface, the silicon film can slip with respect to the current collector during lithiation and delithiation process while maintaining electrical contact.

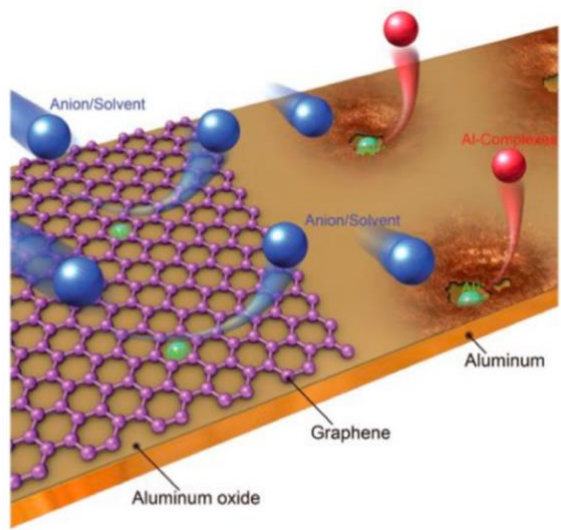

**Figure 13:** Schematic illustration of graphene-armored aluminum current-collector foil with enhanced anti-corrosion property as current collectors for Li-ion batteries. Reprinted with permission from ref[81]. Copyright 2017 @Wiley.

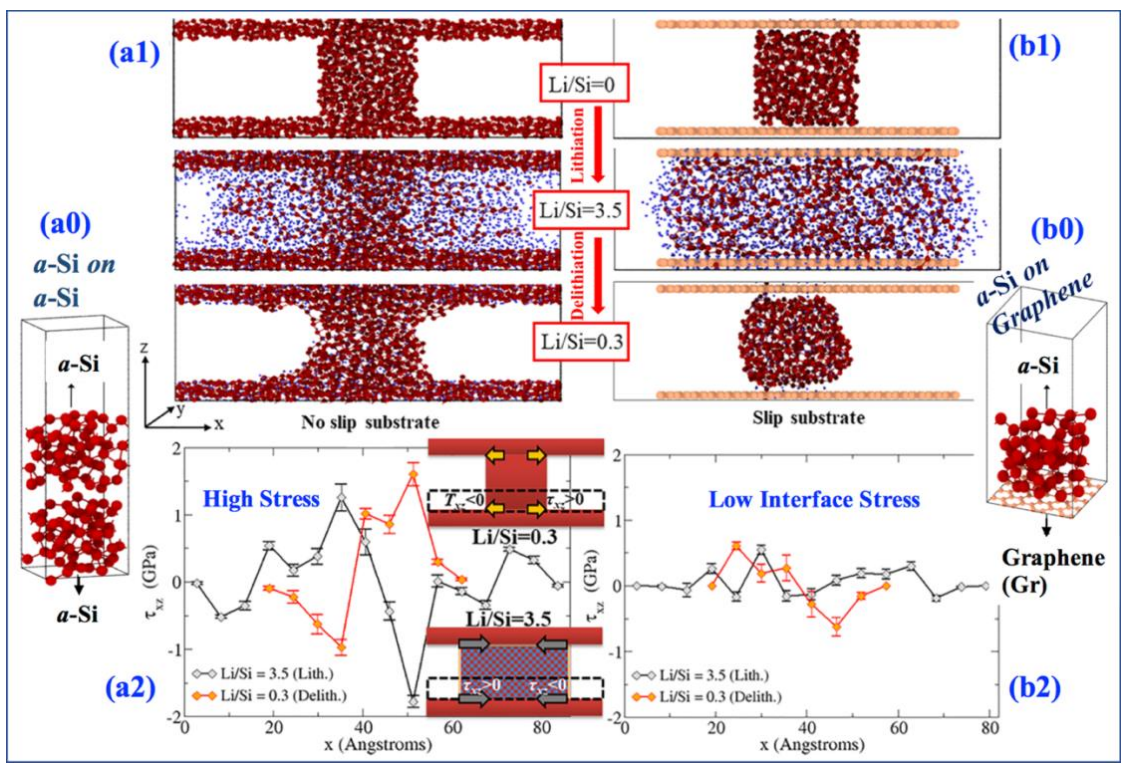

**Figure 14: [a0]** Silicon on silicon substrate. **[a1, b1]** a1 and b1 show snapshots of the simulation systems with a rigid nonslip substrate (a1) and a rigid slip substrate (b1) during lithiation and delithiation cycle. The snapshots are taken prior to lithiation, at a highly lithiated stage (Li/Si = 3.5) and a highly delithiated stage (Li/Si = 0.3), from top to bottom. The red atoms are silicon, while the blue atoms are lithium. **[a2,**

**b2]** shear stress $\tau_{xz}$ profile at Li/Si = 3.5 (black lines) and Li/Si = 0.3 (orange lines) for the systems with nonslip (a2) and slip substrate (b2), respectively. $\tau_{xz}$ is averaged over the region within 1 nm above the bottom substrate (as indicated by the box) every 0.55 nm along the x-direction. The error bar shows the stress fluctuation over 10 independent stress measurements. Reprinted with permission from ref[44]. Copyright 2018 @American Chemical Society.

Molecular Dynamics (MD) simulations support this concept, indicating that a vdW slippery substrate results in (i) less stress build-up and (ii) reduced stress "cycling" compared to a fixed surface. Figure 14 shows MD simulation[44] of lithiation and delithiation of $a$-Si for two cases: $a$-Si/$a$-Si (Figure 14a0) and $a$-Si/graphene (Figure 14b0). Figures 14a1 and 14b1 show lithiation and delithiation for these two cases. Figure 14a2 and b2 show the shear stress ($\tau_{xx}$) profile at lithiation (Li/Si = 3.5, black line) and delithiation (Li/Si = 0.3, orange line) for the system with nonslip (Figure 14a0) and slip (Figure 14b0) surface. For graphene slippery surface (Figure 14b0), lithiated $a$-Si does not stick to the surface (Figure 14b1). Hence the stress at the interface is reduced (Figure 14b2). This innovative approach of utilizing graphene as a vdW slippery interface on current collectors offers the potential for improved stability and performance in silicon-based batteries, mitigating issues related to stress and delamination, and ultimately enhancing battery reliability[44].

### 3.6 Applications of 2D Materials in Batteries Beyond Electrodes

2D materials offer versatility in various aspects of battery technology, extending beyond their use as electrode materials[84, 85]. Here are some notable applications of 2D materials in different battery components: *(1) Electrolyte-Electrode Interface Control:* Ultrathin 2D materials can serve as stable interfacial layers to control the electrode-electrolyte interface. These layers can suppress dendritic growth and uncontrolled SEI formation[86], contributing to safer and more stable battery operation. *(2) Separators:* Separators in LIBs are essential components that prevent electrical contact between the cathode and anode while allowing lithium-ion diffusion. 2D materials, with their mechanical robustness and ion diffusivity, have shown promise as separator materials[87]. *(3) Electrolytes:* Some 2D materials can be utilized as components of the electrolyte in batteries. While traditional electrolytes are often liquid or gel-based, 2D materials can offer unique properties and enhance the performance of the electrolyte in specific battery systems.

### 3.7 Heterostructure Architecture of 2D Materials for Energy Storage

Heterostructure architectures involving 2D materials (Figure 15) offer several advantages and opportunities for enhancing energy storage systems. Some key points regarding 2D heterostructures are:

1. *Advantages of Heterostructure Architectures:* Individual 2D materials have specific properties that may be beneficial for energy storage, but they may also lack other desirable properties. Combining different 2D materials in heterostructures allows for the synergistic utilization of their strengths while mitigating their weaknesses.

2. *Enhanced Specific Surface Area and Ion Transfer:* Heterostructures made from 2D materials typically possess a large specific surface area, which enhances their contact with the electrolyte. This increased contact area improves the kinetics of ion transfer within the electrode material.

However, it's important to manage electrolyte consumption during the formation of a SEI to prevent irreversible capacity loss.

3. *Control of Interactions and Assembly:* The strength of interactions between different layers in heterostructures can be controlled through chemical modifications and assembly parameters. For instance, weak interactions between layers, such as graphene and MXenes, can facilitate ion intercalation and faster ion diffusion compared to interfaces between layers of the same material.

4. *Influence on Electronic Properties:* Direct contact between layers in heterostructures can significantly impact electronic properties. The introduction of stabilizing species, like ions and molecules, at heterointerfaces can enhance electrochemical stability and ion diffusion within the structure.

5. *Exploring Different Interface Architectures:* Researchers should investigate various interface architectures, including both intimate (direct) contact between layers and quasi-intimate interfaces facilitated by through-interlayer-species. These explorations will help determine the most efficient stacked electrode architecture in 2D heterostructures for specific energy storage applications.

Overall, the strategic combination of different 2D materials in heterostructure architectures holds promise for optimizing energy storage performance by leveraging the unique properties of individual components while addressing their limitations.

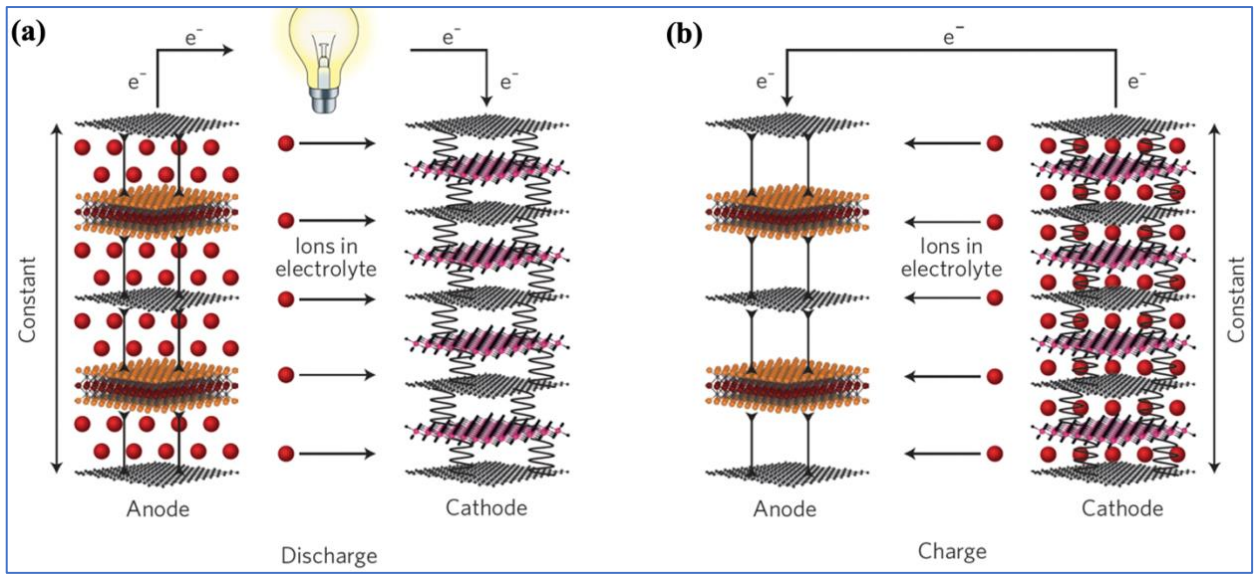

**Figure 15: Schematic illustration of the electrochemical cycling process in a battery with 2D heterostructured pillar electrodes: [a,b]** Battery discharge cycle (a) and charge cycle (b). Pillaring can be used as an effective strategy to conserve the interlayer distance in 2D heterostructures enabling zero-volume change in both electrodes during battery cycling, thus improving mechanical and electrochemical stability of the system leading to extended cycle life. The pillars are schematically represented by elements, shown in black, connecting layers in the 2D heterostructured anode and cathode. Reprinted with permission from ref[63]. Copyright 2017 @ Nature.

### 3.8 Nanofluid Confined 2D Materials for Energy Storage

In the quest to develop energy storage with both high power and high energy densities, while maintaining high volumetric capacity, recent results show that a variety of 2D and layered materials exhibit rapid kinetics of ion transport by the incorporation of nanoconfined fluids[88]. While fast electron transfer can be obtained by a variety of methods, including the addition of highly conductive and high-surface-area carbon, or using metallically conductive active materials, obtaining fast ion transport has proved more challenging. The primary technique to improve ion transport of EESS has been to increase the surface area and particle size to decrease the ion-diffusion distance. However, this comes at the expense of the volumetric energy density and typically leads to increased side reactions. Recent studies highlight the foundational research and emerging strategies to characterize and improve ion transport in EESS based upon intercalation reactions using confined fluids in layered and two-dimensional (2D) densely packed solid-state materials.

While vanadium oxides are representative of the numerous layered materials that are likely to incorporate interlayer solvent molecules upon exposures to a liquid environment, some materials can be synthesized with more strongly bound structural water. Examples include layered and hydrated tungsten and molybdenum oxides: $WO_3 \cdot 2H_2O$, $WO_3 \cdot H_2O$, $MoO_3 \cdot 2H_2O$, and $MoO_3 \cdot H_2O$. These crystalline materials exhibit both secondary-bound interlayer water as well as covalently bound structural water that forms $MO_5(OH_2)$ octahedra (M = W or Mo). Recent studies show that confined fluid molecules lead to improvement in energy storage kinetics in layered materials other than just oxides. In some cases, there is even a transition from battery to capacitor response without increasing the material surface area: a good indication that materials with nanoconfined fluids may be able to exhibit volumetric energy densities that approach those of bulk materials *due to a decrease in volume of micropores and more complete utilization of active materials.*

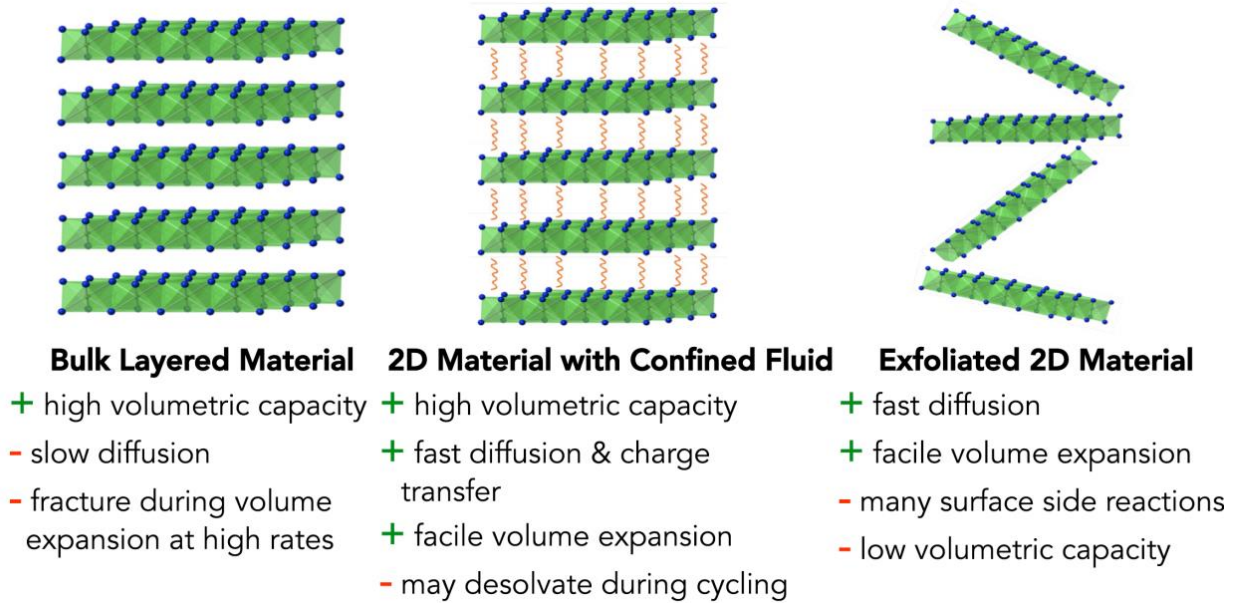

**Figure 16:** Benefits and Drawbacks to Three Types of Materials Structures: Bulk layered material, a 2D materials with a confined fluid, and exfoliated (or nanosheet) 2D materials. 2D materials with nanoconfined fluids represent an intermediate material structure for high power and high energy density

## 4. POTENTIAL PROBLEMS WITH MECHANICS OF 2D MATERIAS IN ENERGY STORAGE SYSTEM

### 4.1 Fracture of Monolayer 2D Materials

***Study of Charge/Discharge-Induced Fracture***

The investigation of fracture mechanisms in various 2D materials has been the subject of extensive research, encompassing both experimental and computational studies. Zhang et al.[14] conducted a comprehensive review, with a primary focus on understanding the fracture behavior of graphene. Most of these studies traditionally involve applying external forces or displacements to induce fractures in the materials. However, when we consider the application of 2D materials in Electrochemical Energy Storage Systems (EESS), such as batteries, fracture mechanisms can manifest due to diverse factors. These factors include ion intercalation/deintercalation, charge/discharge rates, and the influence of the electrolyte. Unlike conventional fracture studies that predominantly emphasize the role of external forces, these fracture mechanisms within EESS are intricately connected to the complex electrochemical processes occurring during energy storage and release.

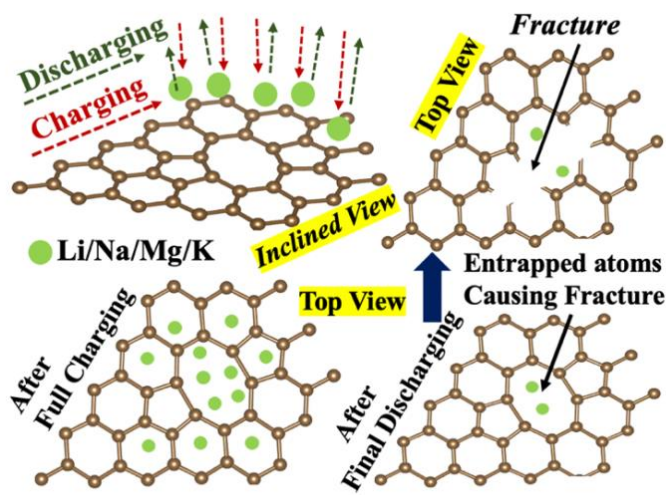

**Figure 17:** Adatoms entrapment/detachment during charging/discharging on defective graphene. Same analogy can be applied to other 2D materials.

For instance, as illustrated in Figure 17, defects in graphene can facilitate the adsorption of adatoms. Similarly, the figure demonstrates that during the charging process, adatoms undergo adsorption. Subsequently, during discharging, the adsorbed atoms become detached and migrate towards the cathode. Importantly, if the defect sites exhibit a propensity for strong adsorption, the discharging process might induce fracture, leading to structural damage. Gaining a deep understanding of these fracture mechanisms is of paramount importance for the development and optimization of 2D materials in electrochemical applications. By unraveling the factors that influence fractures within EESS, researchers can devise more robust and durable materials, thereby enhancing the overall performance and lifespan of batteries and other energy storage devices. Further exploration of the interplay between electrochemical processes and material fractures will undoubtedly pave the way for more efficient and reliable energy storage technologies.

To the best of our knowledge, there is no existing study on fractures in 2D materials that accurately mimics the charge/discharge processes in EESS. Here, we propose a couple of approaches to address this issue:

### (i) Grand Canonical Monte Carlo (GCMC) Simulation of Charge/Discharge on Defective Graphene:

Through GCMC simulations with varying flux rates, adatoms such as Li, Ca, Mg, and Al can be inserted or removed to replicate the charge/discharge processes.

### (ii) Traditional Approach: Applying Strain/Displacement at the Boundary Considering Intercalated 2D Materials:

As shown in Figure 18, this method involves the lithiation of graphene. After determining the adsorption strength ($E_{ad}$) of adatoms on graphene, this information can be utilized to conduct Kinetic Monte Carlo (KMC) simulations to obtain the 'optimal pattern' of lithium distribution over the graphene. Subsequently, external strain or force can be applied to the lithiated graphene. This approach can be extended to other 2D materials and intercalation ions (Na, Ca, Mg, K, etc.).

Apart from intercalation, it's essential to investigate the detailed impact of electrolytes. A comprehensive analysis is needed to understand how the mechanical strength of 2D materials deteriorates upon contact with different electrolytes. Furthermore, the type of electrolyte (organic or aqueous) may exert distinct influences on the fracture properties of 2D materials.

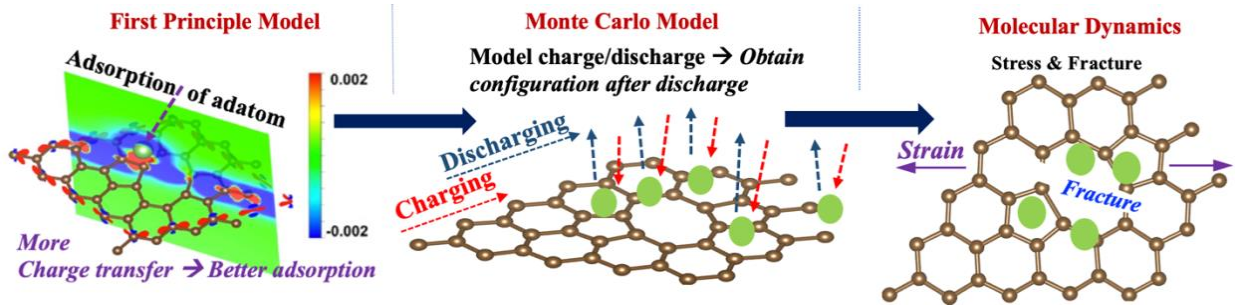

**Figure 18:** A possible approach to study charge/discharge-induced fracture.

The failure strength of a large, thin, brittle plate containing a crack can be expressed as follows[14]:

$$\sigma^{th} = \sqrt{\frac{\Gamma E}{\pi a}} \tag{1}$$

In this equation, $E$ and '$a$' represent the Young's modulus and half of the crack length, respectively. 'Γ' signifies the fracture toughness, which represents the elastic energy released per unit area of crack advancement. Besides crack length, it's essential to consider the effects of intercalation concentration, charge-discharge rate, and electrolyte impact. For instance, intercalated graphene can be viewed as a system where '$E$' for that specific system must be initially computed. Moreover, 'Γ' needs to be determined for various intercalated systems. However, calculating such effective '$E$' and 'Γ' in MD for different 2D materials can be challenging due to the need for suitable interatomic potentials, which are currently unavailable for most systems.

**Strain Rate Effects on 2D Materials' Strength and Modification for Multifactorial Considerations**

Drawing on the Arrhenius formula, which describes the lifetime of a bond under tensile stress, Zhao et al.[89] developed an equation to express the strength of 2D materials as a function of strain rate ($\dot{\varepsilon}$) and temperature (T):

$$\sigma_r(\dot{\varepsilon}, T) = \frac{U_0}{\gamma} + \frac{k}{\gamma} ln\left(\frac{\gamma K \dot{\varepsilon} \tau_0}{n_s kT}\right) T \tag{2}$$

In this equation, '$K$' represents the Young's modulus, '$\tau_0$' is the period of atomic vibration, '$n_s$' denotes the number of sites available for state transition, '$U_0$' stands for the interatomic bond dissociation energy, '$\gamma$' represents the activation volume (adjusted for local stress conditions), and '$k$' is Boltzmann's constant.

However, when we aim to consider the influence of various intercalation/deintercalation ions, charge/discharge rates, and different electrolytes on the strength of 2D materials, we must adapt the governing equation. By incorporating these additional factors, the revised expression becomes:

$$\sigma_r(\dot{\varepsilon}, T, ion\ type, rate, electrolyte) = \frac{U_0}{\gamma} + \frac{k}{\gamma} ln\left(\frac{\gamma K \dot{\varepsilon} \tau_0}{n_s kT}\right) T + \alpha\ (\text{ion}) + \beta(\text{rate}) + \eta(\text{electrolyte}). \tag{3}$$

Here, $\alpha$ (ion) accounts for the specific influence of intercalation/deintercalation ions on the strength of 2D materials, $\beta$(rate) considers the impact of charge/discharge rates on material strength, and $\eta$(electrolyte) factors in the contribution of various electrolytes.

To calculate the strength of 2D materials under these multifactorial conditions, it is imperative to quantify the effects of each added parameter. This involves understanding how different ions affect the interatomic bond dissociated energy ($U_0$), how charge/discharge rates modify the activation volume ($\gamma$), and how various electrolytes influence both the Young's modulus (K) and the period of atomic vibration ($\tau_0$).

It's worth noting that incorporating these additional variables may necessitate a combination of experimental data and theoretical modeling. Accurately determining the relationship between these factors and the strength of 2D materials is a complex task, but it holds the potential to provide a more comprehensive understanding of 2D materials' behavior across a wide range of conditions. Such insights can have far-reaching implications in diverse fields, including materials science and nanotechnology, enabling the development of advanced materials tailored for specific applications and performance requirements.

### Intercalation-Induced Crack: Linear or Nonlinear?

When delving into phenomenon of crack propagation in 2D materials, it's commonly assumed that nonlinear deformation primarily occurs in the vicinity of the crack tip. Consequently, researchers often opt for the simplicity of linear elasticity to analyze crack-related issues in 2D materials. However, recent research has highlighted the significance of nonlinear effects, especially when dealing with very short cracks. A noteworthy study has revealed that the classical Griffith criterion, which is rooted in linear elastic fracture mechanics, tends to overestimate the failure strength of 2D materials when the crack length decreases to below 10 nm. Several crucial questions emerge from this perspective:

1. *How do the processes of intercalation and the rate of charge-discharge affect the linearity or nonlinearity of crack-related problems?*

Interactions between intercalation processes and charge-discharge dynamics can significantly impact the behavior of cracks in 2D materials. The rate at which ions intercalate and deintercalate, as well as the corresponding strain and stress distributions, may lead to nonlinear response in crack propagation. Understanding the interplay between these factors is vital for predicting crack behavior accurately.

2. *What role do electrolyte play in governing these crack-related issues, and how does the type of electrolyte (organic and aqueous) influence the outcome?*

Electrolytes can have a profound impact on the mechanical behavior of 2D materials. The choice between organic and aqueous electrolytes can introduce variations in the material's mechanical properties, which, in turn, affect crack behavior. Electrolyte-specific interactions with the 2D material surface and their ability to influence stress distribution and ion mobility at the crack tip can contribute to nonlinear effects.

Exploring these questions holds the potential to provide insights into the intricate interplay between intercalation processes, charge-discharge dynamics, and the choice of electrolyte type. These factors collectively influence whether the response of 2D materials to cracking follows linear or nonlinear patterns. A comprehensive understanding of these interactions is essential for designing and optimizing 2D materials for various applications, including energy storage systems and structural materials, where crack propagation can have a significant impact on performance and reliability.

### Edge Reconstruction Near a Crack Tip in 2D Materials: Uncharted Territory

The phenomenon of edge reconstruction near a crack tip in 2D materials has garnered attention in atomistic simulations. Surprisingly, even in the absence of intercalation, the effects of such localized changes in atomic structure near a crack tip remain only partially understood. This intriguing area calls for further investigation in the future. The complexity of the problem increases manifold when we consider the influence of factors such as intercalation concentration, charge/discharge rates, and various electrolytes. Remarkably, there is currently a dearth of studies addressing these intricate aspects.

In the realm of MD simulations, a fundamental challenge lies in identifying the appropriate interatomic potential. While the ReaxFF potential serves well for modeling lithiated systems, extending this capability to study other ions (such as Na, Ca, Mg, Al) and a broader range of 2D materials (including Transition Metal Dichalcogenides, or TMDs, and MXenes) necessitates the development of tailored potentials. Recently, there has been an exciting development involving neural network-based potentials that demonstrate the ability to capture phase transitions, such as the transformation from graphite to diamond. This marks a promising direction in the application of modern machine learning techniques to create accurate interatomic potentials for complex atomic systems. Furthermore, molecular mechanics models, replying on classical force fields governing bond, angle, and dihedral interactions, have also been employed to investigate the fracture behavior of graphene. These models offer valuable insights into the mechanical responses of graphene under various conditions.

In summary, the edge reconstruction near crack tips in graphene and other 2D materials presents a captivating avenue for research, warranting in-depth exploration. As we delve into the intricacies of intercalation, charge dynamics, and electrolyte effects, there is an evident need for comprehensive studies that integrate these factors. By harnessing cutting-edge techniques such as machine learning-based potentials and molecular mechanics modeling, researchers are poised to unlock deeper insights into the behavior of 2D materials under the influence of cracks and external factors.

### *Charge/Discharge-Induced Fracture in 2D Materials: Elastic or Plastic Failure?*

It is well-establish that 2D materials like graphene typically exhibit brittle failure. MD simulations have shown that graphene with a sufficiently high density of vacancies can deviate from this behavior and fail in a plastic manner[90]. However, the question of whether intercalated 2D materials, with controlled topological arrangements resulting from processes like intercalation, will fail elastically or plastically remains a subject that requires in-depth investigation, and currently, there is a lack of comprehensive studies addressing this issue. To resolve this question, future research efforts should focus on conducting systematic studies that consider the influence of intercalation, charge/discharge rates, and other relevant factors on the fracture behavior of intercalated 2D materials. Combining experimental investigations with advanced computational techniques may offer insights into whether these materials tend to fail elastically or plastically under different conditions. This knowledge can guide the design and optimization of intercalated 2D materials for specific applications, ensuring their suitability and longevity in practical settings.

## 4.2 Interface Mechanics of Intercalated 2D Materials

### *Interface Adhesion and Correlation with Charge Transfer and Potential Gradient*

In Figure 14, we observe that a graphene-coated current collector serves as a slippery surface, effectively reducing the mechanical stress at the interface. To delve into the fundamental reasons behind this phenomenon, Basu et al.[44] conducted an examination of interface adhesion, specifically focusing on the work of separation ($W_{sep}$) for various interfaces, such as amorphous silicon (*a*-Si) and substrate combinations (Figure 19a). For the *a*-Si/graphene system, the $W_{sep}$ is approximately one-fifth of that observed for Cu and Ni substrates and less than half of that for the *a*-Si/*a*-Si substrate. These findings indicate that graphene provides a slippery interface with *a*-Si, contributing to reduced adhesion forces. In Figure 19b, the correlation between interface strength and charge transfer (Figure 19b1) and potential energy gradient (Figure 19b2) is illustrated. Higher charge transfer and potential gradient values signify strong binding and adhesion at the interface.

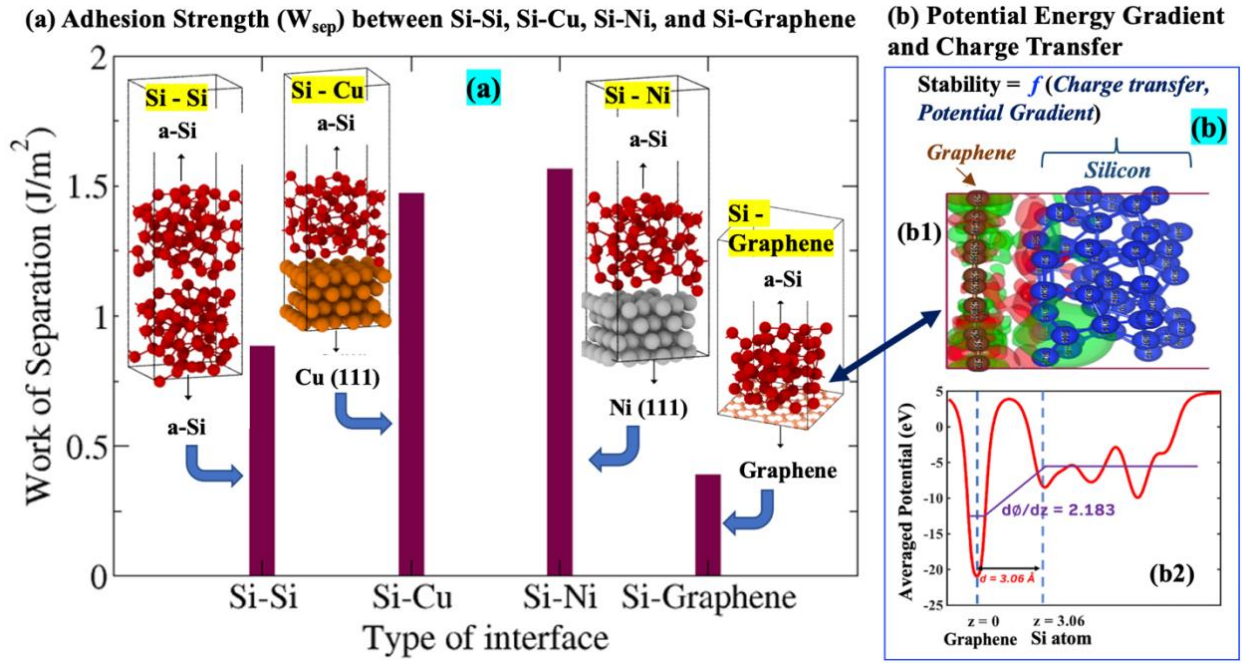

**Figure 19: Work of Separation (J/m²), Charge Transfer, and Potential Gradient. [a]** Work of Separation ($W_{sep}$) for various 3D-3D and 3D-2D interfaces. **[b]** Interfacial charge transfer and potential gradient. Reprinted with permission from ref [31, 44].

In Figure 19b2, the red curve represents the averaged electrostatic potential in the *x-y* plane, while the purple curve represents the averaged potential across the *z* dimension. Across all interfaces, there is a potential step that results in potential gradients ($d\phi/dz$). Here, '*d*' denotes the distance of the nearest Se/Si atom with respect to the graphene sheet. Figure 19b1 presents the charge accumulation (red) and depletion (green). In comparison to Se-graphene interfaces, the *a*-Si/graphene interface exhibits a reduced potential step ($d\phi/dz = 2.18 \ eV/\text{Å}$) and significant overlap of electron clouds across the interface, indicating a higher level of interfacial interaction. Conversely, a large potential gradient ($d\phi/dz$) at Se-graphene interfaces suggests an incohesive interface with limited scope for electron exchange and bonding.

Sharma et al.[31] conducted a study on the ease of electron exchange at silicon-graphene interfaces using the potential gradient and charge separation analysis. A lower potential gradient signifies easier interaction at the interface ($d\phi/dz \sim 0$ for the same materials), while a large potential step suggests an incohesive interface with limited potential for electron exchange and bonding. Charge density in the interfacial region was visualized through charge separation analysis. The charge density plot in Figure 19b1 provides insight into the extent of interaction between atomic systems, consistent with the $\Delta q$ and $d\phi/dz$ results.

It's important to note that this study by Sharma et al.[31] is among the pioneers in investigating the interfacial potential gradient and charge separation at 3D/2D interfaces. However, further in-depth analyses are needed to establish correlations between the interfacial potential gradient, charge transfer, and interface adhesion. Currently, there are no studies that consider the influence of intercalation types (mono- and multivalent ions), charge/discharge rates, and electrolytes on interfacial characteristics. These aspects remain unexplored territory and offer exciting avenues for future research in understanding and optimizing intercalated 2D materials for various applications.

### *Reactions at 3D/3D and 3D/2D Interfaces: Implications and Fundamental Questions*

Figure 19 highlights a critical observation: the 3D/2D interface exhibits less adhesion compared to 3D/3D interfaces. This intriguing mechanical behavior prompts a fundamental question: what is the underlying reason behind this difference? To address this question, Sharma et al.[31] conducted detailed first-principle analyses focusing on Se/graphene and Se/aluminum interfaces. Figure 20 provides valuable insight into the chemical structures of interfaces between *a*-Se/graphene, *c*-Se/graphene, and *c*-Se/aluminum. At the 3D/3D interface, *c*-Se reacts with the aluminum substrate at the interface, leading to some Se atoms becoming attached to the Al surface. In contrast, for the 2D/3D case, no such reaction occurs at the Se/graphene interface. Consequently, the 2D/3D interface yields less adhesion compared to the 3D/3D interface. Moreover, the reaction at the 3D/3D interface generates higher interfacial stress. In the context of EESS applications, this reaction has implications for the availability of "useful" Se atoms for the intercalation purpose. In such cases, the 2D/3D interface offers better performance for EESS. However, several important questions arise:

1. *How will charge/discharge rates influence interfacial characteristics?*

The rate at which charge, and discharge processes occur can significantly impact the behavior of the interface. Faster charge/discharge rates may introduce additional stresses and reactions at the interface, affecting its mechanical and chemical properties.

2. *How do different intercalation ion types (e.g., Li, Na, Ca, Mg, Al, etc.) influence interfacial reactions?*

Different ions can have distinct interactions with the interface and may lead to variations in chemical reactions and adhesion. Understanding the influence of various intercalation ions is crucial for designing materials tailored to specific EESS applications.

3. *What role do different electrolyte types (organic and inorganic) play in influencing interfacial characteristics?*

Electrolyte can have a significant impact on the behavior of interfaces, as they mediate ion transport and reactions. The choice between organic and inorganic electrolytes can affect the interfacial chemistry and mechanics.

4. *Will intercalation influence interfacial reactions, such as in the case of c-Se and aluminum?*

The process of intercalation itself may alter the interface's chemical properties. Understanding how intercalation affects interfacial reactions is essential for predicting the long-term behavior of these interfaces.

Addressing these questions requires a multidisciplinary approach that combines experimental investigations, theoretical modeling, and advanced simulations. Gaining insights into the complex interplay between intercalation processes, charge/discharge rates, intercalation ion types, electrolyte types, and interfacial reactions is essential for optimizing materials for EESS and other applications where interface behavior plays a critical role.

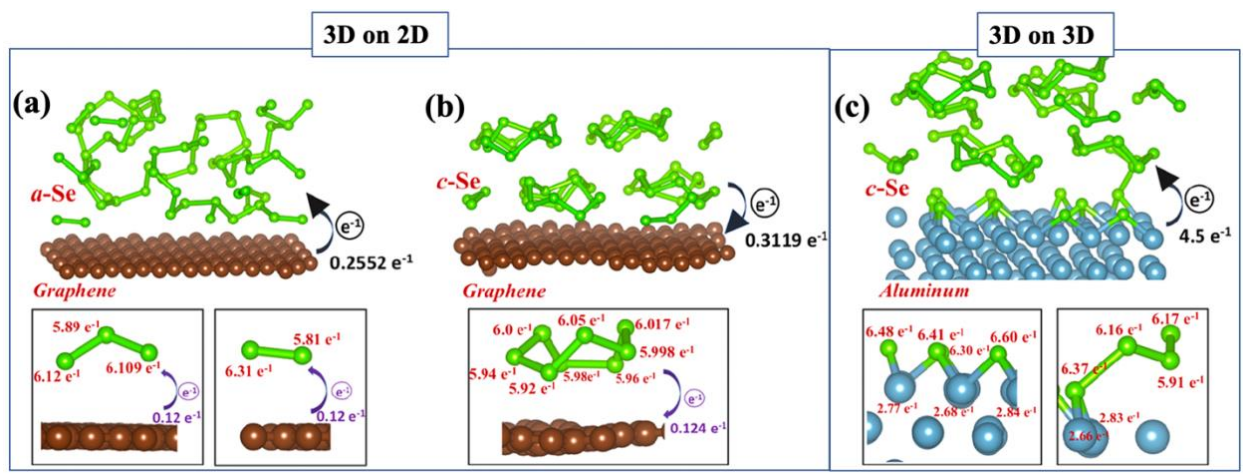

**Figure 20:** Reaction at the interface for **[a, b]** 3D on 2D and **[c]** 3D on 3D cases. **[a]** Illustration of net charge transfer ($\Delta q = 0.2552\ e^{-1}$) from the graphene surface to $a$-Se at the $a$-Se/graphene interface. Se atoms detached from Se chains and attached to fewer than 2-3 Se atoms adsorb on the graphene by gaining more electrons (~0.12 $e^{-1}$). **[b]** The net charge transfer ($\Delta q = 0.3119\ e^{-1}$) at the $c$-Se/graphene interface is directed toward graphene. Se atoms within $Se_8$ rings in the interfacial region have a lower number of electrons than Se atoms farther from the graphene surface. **[c]** Optimized view of the $c$-Se/Al interface where $Se_8$ rings at the interface break into individual atoms to form covalent bonds with surface Al atoms. This surface reaction between Al and Se results in comparatively high net charge transfer ($\Delta q = 4.5\ e^{-1}$) between Se and Al substrate. All atomic charges were obtained via Bader charge analysis. Reprinted with permission from ref[31]. Copyright 2021 @ American Chemical Society.

### *Variation of Interface Adhesion with Intercalation*

An essential question in the study of intercalated 2D materials is whether interface adhesion varies with intercalation. Sharma et al.[31] conducted a study to explore the variation in interface adhesion (measured by $W_{sep}$) with lithiation. The results indicated that $W_{sep}$ does not significantly vary with lithiation. Figure 21 shows that $W_{sep}$ for 50% lithiated silicon ($Li_{0.50}Si/Gr$) is 0.38 J/m$^2$, while for non-lithiated case, $W_{sep}$ is 0.41 J/m$^2$. This simplistic study suggests that lithiation has an insignificant effect on interface adhesion in this specific context.

However, it is crucial to recognize that this observation might not apply universally. Further in-depth analyses are warranted to gain a deeper understanding of how intercalation, especially with different ions such as Na, Mg, etc., influences interface adhesion. Investigating the impact of charge/discharge processes, various intercalation ion types, and their concentrations is vital to comprehend the complexities of interface behavior accurately. Moreover, the effect of intercalation on interface adhesion might be influenced by the specific 2D materials used and their topological characteristics, including defects. Different 2D materials and the presence of defects can lead to variations in interfacial adhesion behavior. Thus, more comprehensive and detailed studies are required to address these nuances and provide a holistic understanding of how intercalation influences interface properties in intercalated 2D material systems.

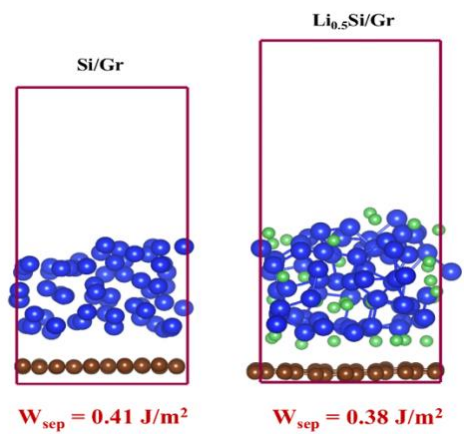

**Figure 21:** Drop in interface strength of graphene with amorphous silicon (Si) electrode as lithium (Li) concentration increased in the system. Interface strength is determined as work of separation ($W_{sep}$) for Si/graphene interface to be 0.41 J/m$^2$ and Li$_{0.5}$Si/graphene to be 0.38 J/m$^2$.

### Crystal Rearrangement at the Interface in 3D/3D and 3D/2D Systems

Sharma et al.[31, 33] investigated how crystal structures at the interface can undergo rearrangement. Figure 22 illustrates this phenomenon, where a distinction is observed between the 3D/3D interface ($c$-Sn$_{64}$ on Cu) and the 3D/2D interface ($c$-Sn$_{64}$ on graphene). For the 3D/3D interface ($c$-Sn$_{64}$ on Cu), there is relatively limited interface lattice rearrangement. In contrast, at the 3D/2D interface ($c$-Sn$_{64}$ on graphene), significant lattice rearrangement of Sn occurs due to the presence of graphene underneath. This lattice arrangement can also be characterized and visualized through the radial distribution function (RDF).

While the specific example in Figure 22 focuses on Sn/graphene, it's imperative to conduct further investigations to understand how different 2D materials and their topological characteristics influence atomic arrangements at 3D/2D interfaces. Each combination of 3D and 2D materials may exhibit unique behaviors and interactions, necessitating a broader exploration of material combinations. Moreover, in the context of EESS applications, understanding how charge/discharge processes affect interatomic arrangement is of paramount importance. The dynamic nature of EESS operations, with repeated charge and discharge cycles, can introduce additional complexities and potentially lead to time-dependent changes in atomic arrangements at interfaces. Consequently, an in-depth investigation into the effect of these processes on interfacial atomic structures is essential for optimizing the performance and reliability of EESS materials and systems.

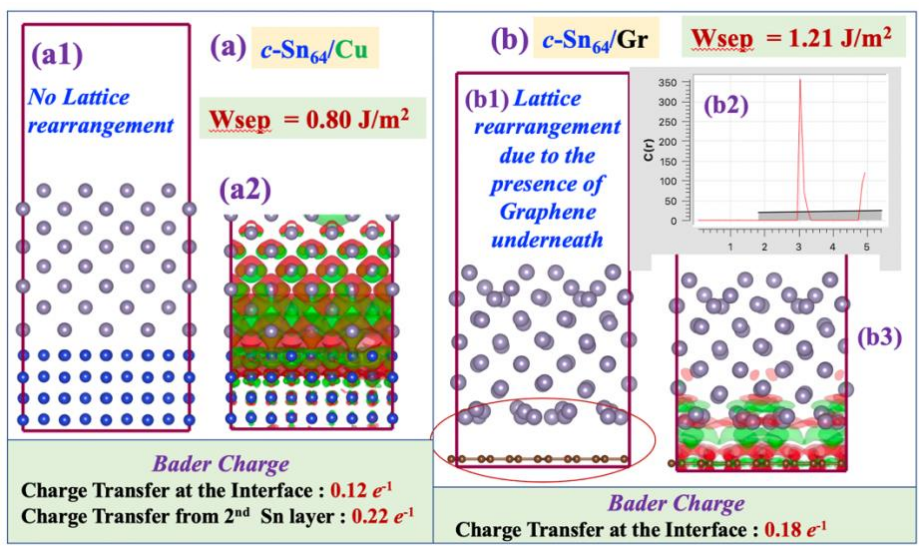

**Figure 22: [a]** No lattice rearrangement at the c-Sn$_{64}$/Cu interface (a1) and interfacial charge transfer (a2). **[b]** Lattice rearrangement at the c-Sn$_{64}$/graphene interface (b1) and associated interfacial charge transfer (b2). Reprinted with permission from ref[31]. Copyright 2020 @American Chemical Society.

***Variation in Interface Adhesion with Surface Termination***

The discussions so far have highlighted that 2D materials often act as slippery interfaces, resulting in reduced interface adhesion. However, it's important to note that different surface terminations of 2D materials can induce variations in interface adhesion. Sharma et al.[30] investigated the interface adhesion variation of MXene with -OH, -OH/O, and -F terminations. The results revealed that MXene with OH-terminated surfaces exhibited high adhesion, while -OH/O and -F terminations drastically reduced adhesion. This indicates that interface adhesion can be tuned by varying the surface functionalization of 2D materials.

Different surface terminations can lead to varying surface reactions, which have implications for interfacial behavior and adhesion. This insight opens avenues for tailoring interface properties by controlling the surface termination of 2D materials. To fully understand the role of surface termination effects during charge/discharge processes, additional research is needed. It's important to investigate how different surface terminations may influence interfacial reactions, ion transport, and adhesion under dynamic electrochemical conditions. Moreover, the effect of electrolyte types, ion types (e.g., Li, Na, Ca, Mg), and charge/discharge rates on interfacial reactions and adhesion is a complex and multifaceted topic that requires in-depth exploration. A comprehensive understanding of these factors is essential for designing materials with tunable interfacial properties for various applications, including energy storage and conversion technologies.

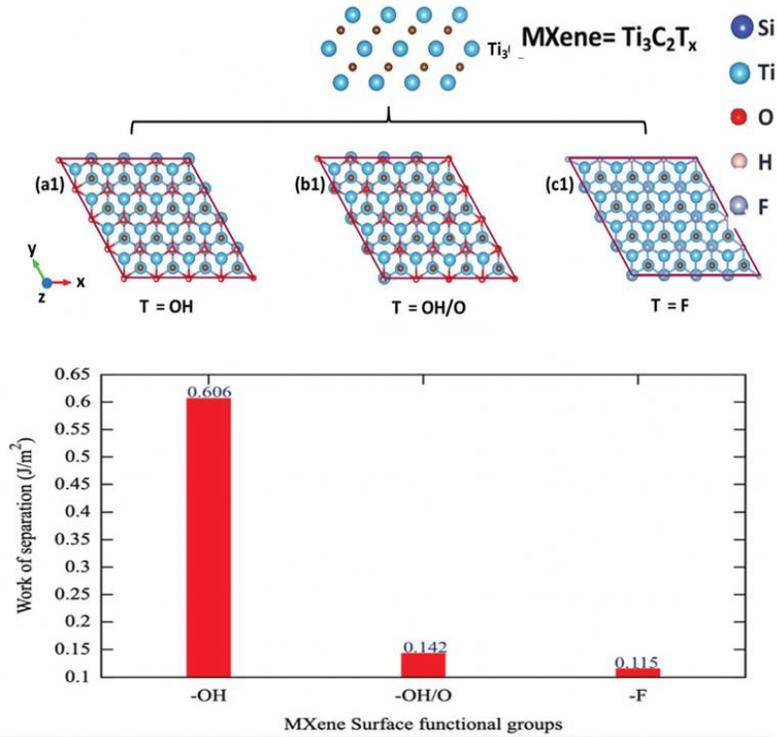

**Figure 23:** Variation of work of separation for different surface functional group terminated MXenes (Ti$_3$C$_2$T$_x$, T = OH, OH/O, F). Reprinted with permission from ref[30]. Copyright 2020 @ Royal Society of Chemistry.

### *Machine Learning for Predicting Interface Properties: A Cost-Effective Approach*

Determining interface properties, as illustrated in Figure 24, can be computationally extremely expensive. To address this challenge, Sharma et al.[33] developed a Modified High Dimensional Neural Network (MHDNN) approach for predicting the properties of interface systems. Instead of relying on a single neural network for the global Potential Energy Surface (PES), this approach describes atomic interactions based on local chemical environments. In this method, the sum of atomic energy ($E_j^i$) contributes to the total potential energy:

$$E_{short} = \sum_{i=1}^{n} \sum_{j=1}^{N_i} E_j^i \qquad (4)$$

Here, 'n' represents the number of elements in the system, and $N_i$ is the number of atoms of element $i$ (Figure 24). Figure 25 illustrates the energy variation of interfaced structures obtained from the Density Functional Theory (DFT) simulations and the machine learning model.

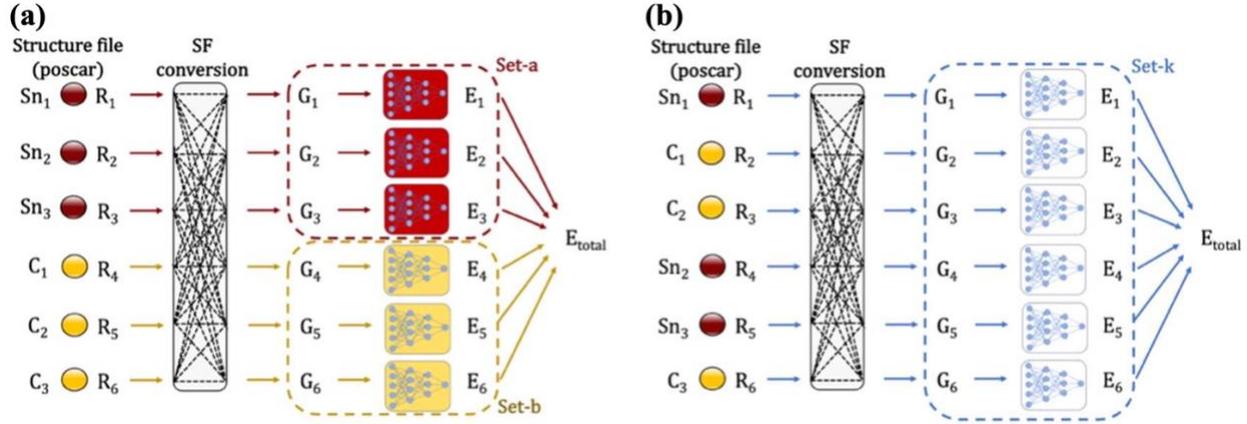

**Figure 24:** Modified High-Dimensional Neural Network (MHDNN): Comparative schematic of HDNN for bicomponent (Sn|C) system: **[a]** HDNN by Behler and Parrinello (BPNN) for bicomponent systems where weights and architecture of atomic neural networks (ann) are the same for single chemical species. Red-ann in set-a corresponds to Sn atoms and yellow-ann in set-b corresponds to C atoms and **[b]** Modified HDNN in the present study for bicomponent systems. Weights and architecture of all atomic neural networks (ann) are the same and correspond to the Sn|C system rather than single species. Atomic species are differentiated by the added feature of atomic number. Reprinted with permission from ref[33]. Copyright 2022 @American Society of Mechanical Engineers (ASME).

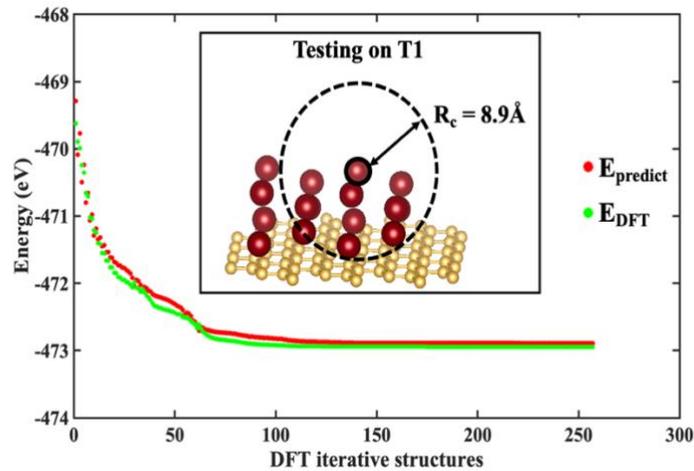

**Figure 25:** Comparison of energy obtained from DFT and MHDNN. Reprinted with permission from ref[33]. Copyright 2022 @American Society of Mechanical Engineers (ASME).

### *Extension of Interface Study to Various Other Systems*

Figures 19-25 collectively demonstrate interface results for specific cases such as silicon/graphene, selenium/graphene, etc. While these results are promising and enable the exploration of interface systems, it's important to acknowledge that performing first-principle calculations for interface systems is computationally demanding and often resource-intensive. Additionally, MD simulations of adhesion

analysis at the atomic scale are not always considered reliable in this context. However, the significance of investigating various interface systems, particularly in the context of energy storage and conversion systems, cannot be overstated. The wide range of 2D materials, including graphene, Transition Metal Dichalcogenides (TMDs), MXenes, and bulk materials like silicon, germanium, and selenium, presents virtually endless possibilities for interface systems. Machine learning approaches like MHDNN offer a cost-effective alternative to expensive first-principle calculations, making it feasible to explore a broader range of interface systems and their properties.

### Interface Stress Analyses

Figure 14 provides valuable insight into the interface stress variation during lithiation and delithiation for silicon on silicon and silicon on graphene systems. These calculations, conducted using ReaxFF and LJ potentials, offer valuable data for understanding mechanical behavior. However, there are significant challenges and areas for further research. *(i) Lack of Interatomic Potentials:* The absence of suitable interatomic potentials for systems like Si/Ni and Si/Cu poses a limitation. To comprehensively understand the interface stress variation during lithiation/delithiation, it is essential to analyze active materials (e.g., Si, Se, Ge) over various current collectors (e.g., Ni, Cu). Developing accurate interatomic potentials for these systems is a critical research need. *(ii) Influence of Different 2D Materials:* Investigating how different 2D materials, including Transition Metal Dichalcogenides (TMDs), MXenes, etc., influence interface stress generation is crucial. *(iii) Complexity of Interactions:* Modeling interface stress in heterogeneous systems involving 3D materials, 2D materials, and current collectors is inherently complex due to the diverse interactions involved. Addressing these challenges for deeper understanding of interfacial stress for various systems is of utmost importance.

### Mechanical Properties of 2D Materials Interfaced with 3D Materials

Studying the mechanical properties of 2D materials when interfaced with 3D materials is a critical aspect of understanding the behavior of these materials in various applications, including energy storage systems. Here are two key considerations:

1. *Stress in 2D Materials:* When 2D materials are used in energy storage applications, they are often placed over a current collector. During the charging/discharging, these 2D materials can undergo stress variations due to the mechanical forces exerted by the electrochemical reactions. Investigating how these stress variations affect the structural integrity and performance of the 2D materials is a topic that requires in-depth investigation. Understanding how stress impacts the 2D materials can help optimize their design for enhanced durability and longevity in energy storage systems.

2. *Topological Changes:* Many studies have explored the phenomenon of wrinkling formation in 2D materials when placed on substrates. For example, substrate strain can induce the formation of wrinkles or other topological changes in 2D materials. When combined with the charge/discharge processes of active materials in energy storage systems, it is essential to investigate how these processes influence the topological characteristics of the 2D materials. Understanding the interplay between mechanical stress, electrochemical reactions, and topological changes is crucial for designing reliable and efficient energy storage materials and systems.

In summary, while 2D materials are often used as a slippery surface in energy storage systems to reduce interface stress, it is equally important to consider the mechanical responses of the 2D materials themselves. This includes studying stress variations in 2D materials during charging/discharging and examining how the combination of current collectors and electrochemical processes can induce topological changes in the 2D materials. Comprehensive research in these areas will contribute to the development of more robust and high-performance energy storage technologies.

### 4.3 Mechanics of 2D Homo- and Heterostructures for Energy Storage

***Interlayer Formation Energy and Charge transfer***

Understanding the stability and interlayer behavior of heterostructures is crucial for their potential applications in energy storage. The example of Tungsten Disulfide ($WS_2$) and Diselenide ($WSe_2$) heterostructures ($WS_2$-$WSe_2$) is considered in Figure 26[22]. Figure 26a, d show the *chalcogen* (*c-*) and *metal* (*m-*) terminated $WS_2$-$WSe_2$. For each termination, AA- (Figure 26b1, e1) and AB-stacking (Figure 26c1, f1) systems are considered.

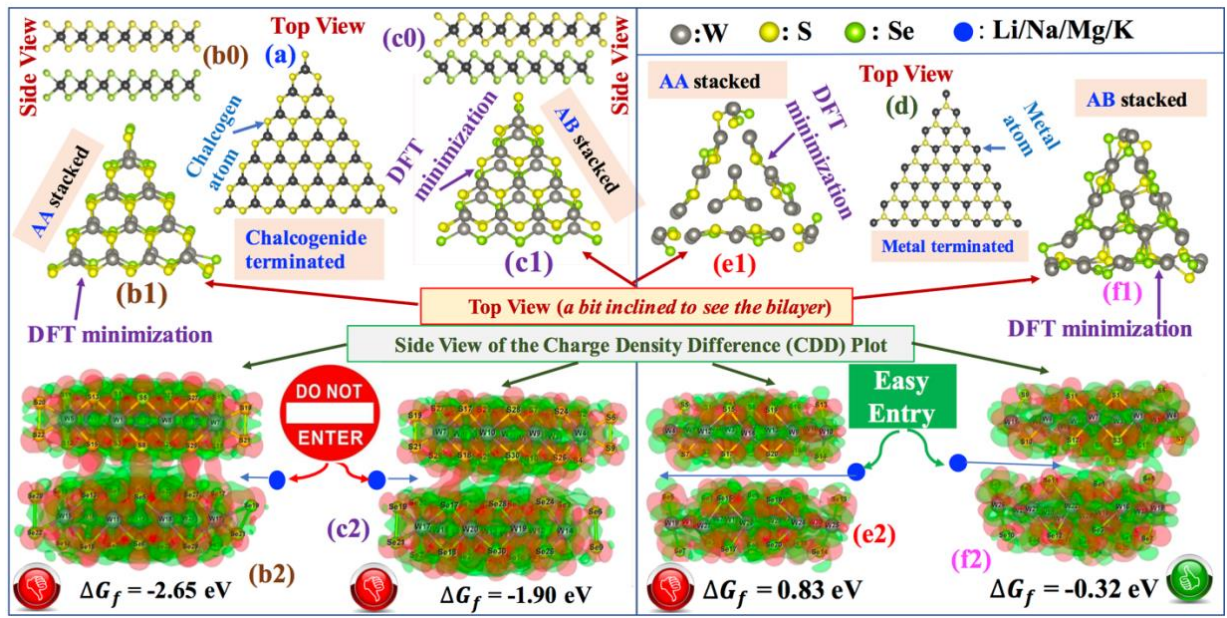

**Figure 26:** DFT study of stability of $WS_2$/$WSe_2$ heterostructures considering various edge terminations and stacking for energy storage. Reprinted with permission from ref[22]. Copyright 2020 @ Nature.

DFT minimization, Gibbs free energy ($\Delta G_f$), and charge-transfer analysis were performed for all four cases. From the DFT minimized structure, we conclude that *c*-terminated AA-stacked (Figure 26b1) is the most stable structure ($\Delta G_f$ = -2.65 eV), while *m*-terminated AB-stacking is the least stable ($\Delta G_f$ = -0.32 eV). AA-stacking for *m*-termination is not stable since $\Delta G_f$ is positive ($\Delta G_f$ = 0.83 eV). Charge Density Difference (CDD) analysis shows that for the most stable structure (Figure 26b2), charge overlaps between layers indicating strong interlayer interaction. Charge overlap reduces as the stability reduces. *For the **energy*** *storage purpose*, fast ion transport and kinetics between the layers of 2D materials is necessary. Charge-overlap acts as "*Do Not Enter*" zone blocking the fast transport (Figure 26b2, c2). No charge overlap corresponds to an unstable structure (Figure 26e2), which can't be used for energy storage. Hence, intermediate stable structure with less charge-transfer is most preferable (Figure 26f2). The concept is analogous to $W_{sep}$ between bulk and 2D materials (Figure 19), where very high or low $W_{sep}$ is not good for energy storage application. Here, configuration in Figure 26f1,f2 is the most 'preferable' as compared to configurations in Figure 26 b1,b2 and c1,c2 with stronger interlayer adhesion ($\Delta G_f$).

The example highlights the importance of carefully analyzing the interlayer behavior and charge transfer in 2D heterostructures before selecting materials and configurations for energy storage applications. Additionally, it emphasizes that the properties of 2D materials and their heterostructures can vary depending on the arrangement, and thorough investigations are required to understand their behavior under different conditions. It's important to note that the analysis presented here focused on the stability and interlayer interactions of 2D heterostructures without considering the effects of charge/discharge processes and intercalation ion types. These factors can further influence the behavior of 2D materials in energy storage applications, making detailed investigations crucial for their successful implementation. In conclusion, understanding how interlayer interactions, intercalation, and charging/discharging processes impact the performance of 2D materials and their heterostructures is essential for their potential use in a wide range of applications, including electronics, optoelectronics, energy storage, and sensors. Thorough research in these areas is key to unlocking the full potential of these materials.

### Interlayer Mechanical Stability and Phase Transition During Charge/Discharge

Understanding the interlayer mechanical stability and phase transitions in 2D materials during the charge and discharge processes is essential for the development of effective energy storage systems. Here, we discuss two studies involving Molybdinum Disulfide (MoS$_2$) as an example to illustrate these phenomena:

1. *Lithiation/De-lithiation of 2H-MoS$_2$:* Zhang et al.[91] observed that the intercalation reaction of 2H-MoS$_2$ is reversible and involves a phase transformation between the 2H phase and a disordered 1T phase. As the intercalation progresses further, the disordered 1T phase leads to structural degradation, eventually resulting in complete structural degradation.

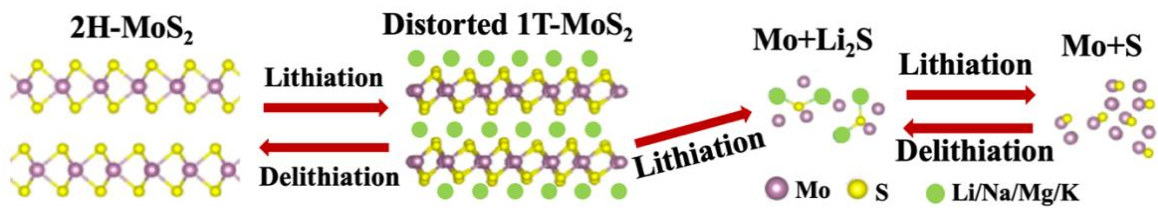

**Figure 27:** Dissolution of polysulfides for cycled MoS$_2$ electrodes. Phase transition of 2H-MoS$_2$ to distorted 1T-MoS$_2$. Subsequent lithiation/delithiation causes structural degradation. Reprinted with permission from ref[91]. Copyright 2018 @American Chemical Society.

2. *Sodium Intercalation in 1H-MoS$_2$:* In another study, He et al.[92] investigated the phase transition of sodium intercalation in 1H-MoS$_2$. They found that sodium intercalation transforms 1H-MoS$_2$ into a distorted 1T-MoS₂ phase. Subsequent intercalation and deintercalation further transform the distorted 1T-MoS₂ into a ZT-MoS₂ phase (Figure 28). The Nudged Elastic Band (NEB) method was used to calculate the energy barrier for the phase transition from 1H to 1T structure. The barrier energy reduced significantly when sodium atoms were adsorbed on one side of the MoS₂ monolayer, suggesting the role of sodium in the phase transition process.

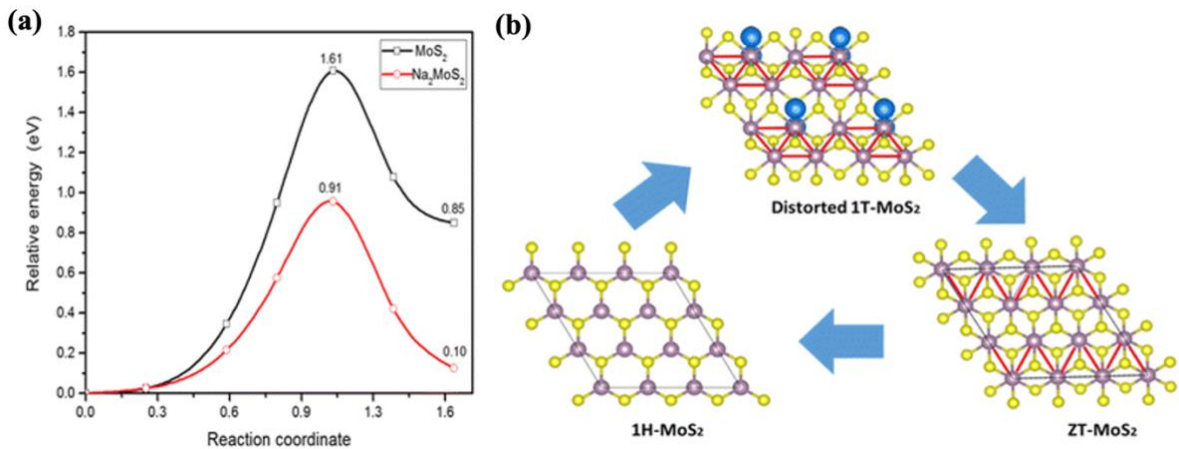

**Figure 28: [a]** Evolution of the energy per S atom for 1H to 1T structure transition as a function of the reaction coordinate, for pure and Na-covered MoS₂. **[b]** The pathways of structural phase transition of Na adsorption on monolayer MoS₂. Reprinted with permission from ref [91]. Copyright 2018 American Chemical Society.

These examples demonstrate that during the charging and discharging of energy storage systems based on 2D materials, two important phenomena can occur:

*(i)* *Complete Structural Degradation:* The intercalation process may lead to complete structural degradation of the 2D material, rendering it unusable for further energy storage.

*(ii)* *Phase Transformation:* The intercalation process can also induce phase transformations in the material, leading to different structural configurations. These phase transitions may be reversible or irreversible depending on the material and the intercalation conditions.

Given these possibilities, it is crucial to computationally investigate interlayer stability and phase transitions in various 2D homo- and heterostructures for different intercalation ions. Additionally, considering that multivalent ions (e.g., Ca, Mg, Zn) are bigger in size than monovalent ions (e.g., Li, Na) and exhibit different chemical behaviors, it is hypothesized that the type of intercalation ion can significantly influence interlayer stability and phase transition. In summary, a comprehensive understanding of interlayer stability and phase transitions in 2D materials during charge/discharge processes is essential for designing reliable and efficient energy storage systems. Computational studies can help protect and optimize the behavior of these materials under various conditions and with different intercalation ions, leading to advancements in energy storage technology.

*Interlayer Friction and Fracture of 2D Materials During Charging/Discharging*

Understanding how interlayer interactions and friction in 2D materials affect crack propagation is essential for designing reliable energy storage systems. Here, we discuss a study involving bilayer molybdenum disulfide ($MoS_2$) to illustrate these phenomena.

*Effect of Stacking on Friction:* The study by Jung et al.[93] found that the magnitude of interlayer friction significantly depends on the stacking arrangement and loading conditions. However, the analysis revealed that the relative positions of sulfur atoms in the top and bottom layers also play a crucial role in determining friction. Friction forces peak when the sulfur atoms in the top layer align with those in the bottom layer.

*Crack Propagation in Bilayer $MoS_2$:* Jung et al.[93] investigated (Figure 29) the dynamics of crack propagation in suspended bilayer $MoS_2$ and explored how the uncracked layer influences crack propagation in the cracked layer. Subsequent MD simulations measured the friction between the two layers with different stacking and loading conditions.

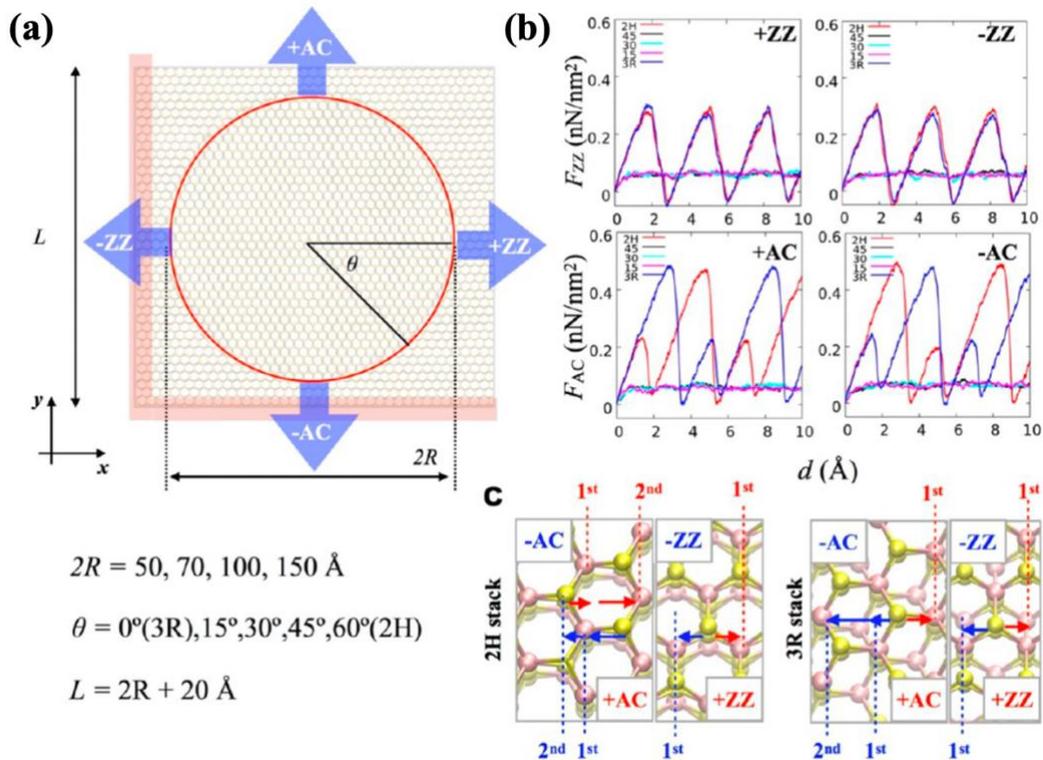

**Figure 29: [a]** Schematic depiction of MD simulations for interlayer frictional forces per area (eV/nm3) with four different loading directions: positive and negative zigzag (+ZZ and -ZZ) and armchair directions (+AC and -AC). **[b]** Forces obtained from the system size of 2R = 150 Å. (c) Relative positions of sulfurs of top and bottom layers. Considering the small difference from weak vdW interactions, the 10 times higher

friction force is counterintuitive. Reproduced with permission from ref[93]. Copyright 2018 American Chemical Society.

*Crack Propagation and Friction:* MD simulations were conducted with a circular bilayer under different stacking conditions. The results showed that:

(i)   When the stacking alignment reduces interlayer friction, the crack can pass through the bottom layer without breaking the top layer (Figure 30e).

(ii)  Coherent fracture occurs when both layers move together during crack propagation (Figure 30f). In this case, the top layer breaks after the crack passes through the bottom layer.

These findings highlight the significance of interlayer orientation in fracture behavior, which is influenced by different frictional characteristics.

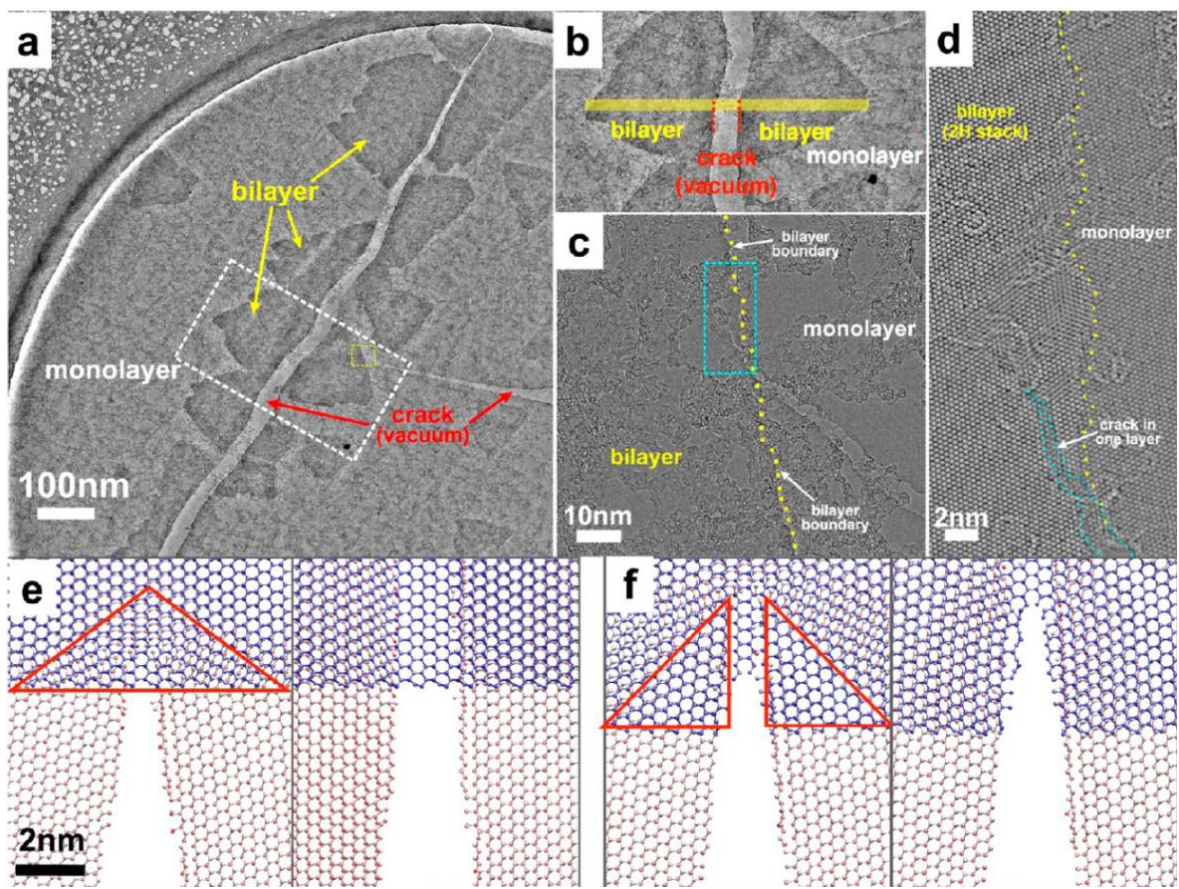

**Figure 30: [a]** Low-magnification TEM image showing two-cracks on a suspended MoS$_2$ region. Most areas of this suspended MoS$_2$ region are monolayer, with some small bilayer/multilayer islands distributed on it, as marked by yellow arrows. Two cracks can be observed. **[b]** Zoom-in TEM image of the region marked by the dashed white box on the left crack in panel a. **[c]** Zoom-in TEM image of the region marked by the dashed yellow box at the tip region on the right crack of panel a. The crack goes from the monolayer MoS2 into the bilayer region. **[d]** Zoom-in TEM image of the region in the cyan dashed box in panel c, confirming

that the crack is restricted to one layer when it goes into the bilayer region. **[e,f]** Snapshots of MD simulations. **[e]** The crack passes through in the bottom layer without breaking the other layer because the interlayer friction is significantly reduced by the disturbed stacking alignment. **[f]** Coherent fracture occurs when both layers move coherently during the crack propagation. The red triangle indicates the regions where the effective tensile stress locally occurs due to the high friction from the 2H stack alignment. The top layer breaks after the crack passes through in the bottom layer. Reprinted with permission from Ref[93]. Copyright 2018 American Chemical Society (ACS).

When applied in energy storage systems, such as batteries, the intercalation of ions inside 2D layers is likely to influence interlayer interactions and fracture behavior. Additionally, the type of intercalation ions (e.g., Li, Na, K, Mg, etc.) will govern interlayer friction and fracture behavior. The presence of defects in 2D materials will also impact friction and fracture patterns. These findings raise important questions about the interplay between interlayer interactions, ion intercalation, and fracture behavior in 2D materials. To address these questions comprehensively, further studies are needed, considering various 2D homo- and heterostructures, intercalation ions, topological defects, electrolyte effects and more. Understanding these complex interactions will be crucial for optimizing energy storage systems and ensuring their long-term reliability and performance.

## 4.4 Mechanical Properties of 2D Materials with Nanofluid Inside During Charging/Discharging

The early studies mentioned in section 3.8 (Figure 16) found that interlayer water could be unstable or removed after electrochemical cycling in highly alkaline or aprotic electrolytes, the concept of utilizing interlayer solvation and the nanoscale confinement of fluids for improving ion transport remained underexplored in the coming decades. There is significant opportunity to understand the detailed mechanisms of energy storage within redox-active materials with nanoconfined fluids. The potential benefits of these materials derive from the *effects of nanoconfined fluids on interfacial charge transfer and ion diffusion*. The potential effect would be a decrease in the charge-transfer activation energy.

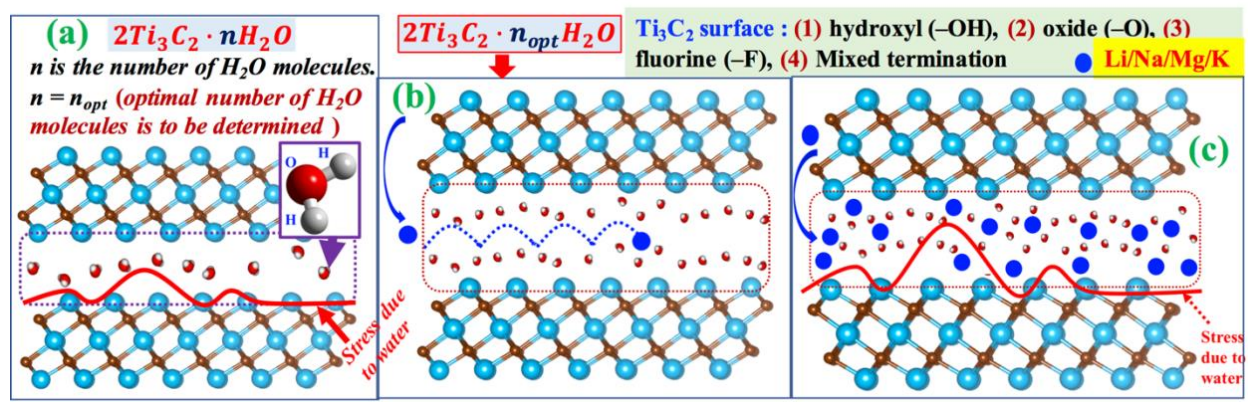

**Figure 31:** Stress generation on 2D MXenes surface by nano-confined fluids.

There are many important open problems that chemo-mechanics community can addresses: The interlayer may be increased to large dimensions ($> 40\text{Å}$) using nanoconfined fluids. At which point do the walls of the materials no longer 'sense' each other? It is important to study interlayer interaction (formation energy) with varying water contents between the layers. Moreover, for a given layer distance, how does the water content inside influence the interlayer formation energy? Is there any optimal water concentration for most stable formation energy and charge transfer calculation?

How do the number of fluid layers and their bonding to the framework affect the mechanical properties? For example, stress generated on 2D materials due to fluid inside layers needs to be investigated. Moreover, this nanofluid-induced stress may vary with charge/discharge rate, intercalation ion types, fluid type, fluid content, various types of 2D materials and their surface terminations.

We discussed in section 4.3 and in Figures 27, 28 about phase transformation in TMDs during charging/discharging. How will nanoconfined fluids affect this phase transformation need in-depth further investigation. Kim et al.[94] studied graphene-water interaction for drug delivery problem using MD simulation. The reliable interatomic potential for graphene-water system is available. However, currently, suitable potential for other water-2D materials is unavailable. Moreover, there is no suitable potential to model systems involving water, intercalation ions, 2D materials. Inclusion of electrolyte makes the situation even more complicated. DFT calculations on these systems are very difficult because of high computational cost involved. Therefore, there is urgent need to develop interatomic potentials for these systems. Machine learning potential for these systems will be very useful and needs a lot of work. As mentioned before, training data for developing machine learning model needs to be completely generated. There is no existing database such as Materials Project[35], OQMD[37] for these water-intercalated systems.

## 4.5 Study of Wrinkling, Folding, Crumping in Intercalated 2D Materials

As shown in Figures 8-12, graphene crumpled, bended to encapsulate electrode materials. The study of wrinkling, folding, and crumpling in intercalated 2D materials, especially under the influence of charge/discharge processes and electrolytes, is an important area of research with potential implications for various applications. Here are some considerations and potential areas for investigation:

*(i) Charge/Discharge-Induced Wrinkling:* Understanding how the intercalation of ions during charge and discharge cycles influences the structural integrity and wrinkling formation in 2D materials is crucial.
*(ii) Electrolyte Influence:* Different types of electrolytes (e.g., aqueous, organic) can interact differently with 2D materials, potentially affecting their mechanical properties and propensity for wrinkling. Investigating the role of electrolyte composition, concentration, and pH in wrinkling formation can provide valuable insights into the behavior of intercalated 2D materials.
*(iii) Ion-Specific Effects:* Different ions used for intercalation (e.g., Li, Na, Mg) may have varying sizes, charges, and chemistries, which can influence how they interact with 2D materials and affect formation of wrinkling, crumpling, folding.
*(iv) Temperature and Rate Dependence:* The temperature and rate of charge/discharge processes can impact the kinetics and ion intercalation/deintercalation. Investigating how temperature and rate variations influence wrinkling behavior is essential for a comprehensive understanding of the material's response.

*(v) Characterization Techniques:* Researchers can employ advanced characterization techniques such as atomic force microscopy (AFM), scanning electron microscopy (SEM), and transmission electron microscopy (TEM) to visualize and quantify wrinkling patterns in intercalated 2D materials. These techniques can provide valuable data on the morphology and extent of wrinkling.

*(vi) Materials Variety:* Investigating a wide range of 2D materials beyond graphene, such as Transition Metal Dichalcogenides (TMDs), MXenes, and other emerging 2D materials, is essential. Different materials may exhibit unique wrinkling behaviors due to variations in their mechanical properties and structural characteristics.

*(vii) Applications and Device Performance:* Understanding how wrinkling affects the performance of energy storage and other applications is crucial. Researchers can explore the role of wrinkling in improving or compromising device efficiency, longevity, and reliability.

In summary, the study of wrinkling, folding, and crumpling in intercalated 2D materials is an exciting and evolving research area with significant potential for impact in various fields, including energy storage, nanoelectronics, and materials science.

### 4.6 Study of Electrolyte-Electrode Interfaces in 2D Materials-Based EESS

The study of the interfacial mechanics between 2D materials electrodes and electrolytes in energy storage systems is crucial for understanding the performance and durability of these materials. Here are some key areas for investigation:

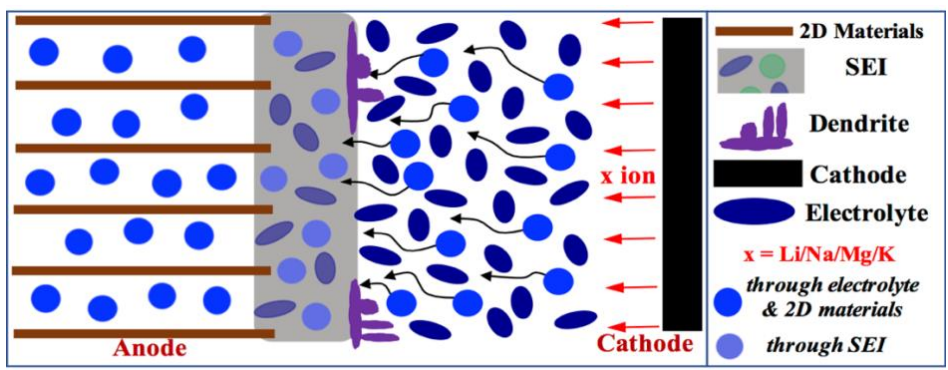

**Figure 32:** Electrode and electrolyte interface in 2D materials-based battery.

*(i) Solid Electrolyte Interphase (SEI) Formation:* SEI is a critical component of many electrochemical energy storage systems, such as lithium-ion batteries. Understanding how SEI forms on the surface of 2D material electrodes and characterizing its mechanical properties is essential.

*(ii) Mechanical Properties of SEI:* SEI is not just a passive layer; it can undergo mechanical stress and deformation during charge and discharge cycles. Investigating the mechanical properties of SEI, including its adhesion to the 2D material electrode and its response to stress, can provide insights into its stability and role in the overall performance of the energy storage systems.

*(iii) Electrolyte-Induced Effects:* Different types of electrolytes (e.g., aqueous, organic) can interact differently with 2D material electrodes. Research should explore how the choice of electrolyte affects the mechanical properties of the electrode-electrolyte interface and the formation of SEI. The composition, concentration, and pH of the electrolyte can all play a role.

*(iv) Dendrite Formation:* Dendrite growth on the electrode surface is a common issue in some energy storage systems, particularly lithium-ion batteries. Investigating the conditions that promote dendrite formation and the mechanical properties of dendrites is crucial for addressing safety concerns and improving the longevity of energy storage devices.

*(v) Materials Compatibility:* Different 2D materials may exhibit varying interactions with different electrolytes. Researchers should explore the compatibility of various 2D materials with different electrolyte chemistries to identify materials that are well-suited for specific energy storage applications.

*(vi) Effect of Charge/Discharge Rate:* The rate at which the charge and discharge processes occur can influence the mechanical behavior of the electrode-electrolyte interface. Investigating how different charge/discharge rates affect SEI formation, dendrite growth, and overall mechanical stability is important.

*(vii) Impact on Device Performance:* Ultimately, the goal is to understand how the mechanical properties of the electrode-electrolyte interface impact the performance, safety, and longevity of energy storage devices. This includes evaluating the influence on capacity, cycling life, and thermal stability.

By studying these aspects of the electrode-electrolyte interface in 2D materials-based energy storage systems, researchers can make important strides in improving the design and performance of next-generation energy storage technologies.

## 5. COMPUTATIONAL CHALLENGES AND FUTURE DIRECTIONS

Addressing the computational challenges in the study of 2D materials and their applications in energy storage is crucial for advancing this field. Here are some potential solutions and future directions:

**Density Functional Theory:**

Density Functional Theory (DFT) serves as the foundation for calculating interlayer formation energy and charge transfer in this study. While MD simulations may seem like an alternative, they are not suitable for accurately determining formation energy and charge transfer. Nevertheless, it's worth noting that DFT calculations are computationally intensive, especially for periodic systems. The non-periodic systems depicted in Figure 26 demand even more computational resources for formation energy and charge transfer calculations. Consequently, attempting DFT calculations for the countless 2D homo- and heterostructures is a computationally infeasible endeavor. However, it's worth mentioning that some DFT codes, such as SPARC[95] and DFT-FE[96], have been developed to handle large systems efficiently and claim the ability to compute millions of atoms within reasonable timeframes. Yet, it's essential to acknowledge that there is no existing research on the application of these codes to the diverse array of structures mentioned in this study. Therefore, investigating the implementation of these advanced DFT codes for these systems could be an intriguing avenue for future research.

**Molecular Dynamics Simulation:**

MD simulations have been employed to investigate various aspects, including interlayer stability, phase transitions resulting from interlayer intercalation, interlayer friction, and fracture behavior. However, when considering intercalation ions like Li, Na, Mg, Ca, and others, the availability of suitable interatomic potentials becomes even scarcer. There has been no prior research examining how intercalation influences interlayer friction and fracture in depth. In MD simulations, having the right interatomic potential is critical. Unfortunately, for most 2D homo- and heterostructures, finding an appropriate interatomic potential remains a challenge. Consequently, it is of utmost importance to expedite the development of suitable MD potentials tailored for studying interlayer chemo-mechanics across various intercalation ions.

**Continuum Modeling:**

Continuum modeling has seen applications in the context of 2D materials[97, 98], but, to the author's knowledge, there have been no studies specifically focused on continuum modeling in the realm of 2D materials based EESS. It's important to recognize that while continuum models can provide valuable insights into macroscopic behaviors and trends, they fall short in capturing the finer atomic-level details that atomistic modeling excels at providing. Given the unique characteristics and applications of 2D materials in the field of EESS, there may be opportunities to develop tailored continuum models that bridge the gap between the atomic scale and macroscopic behavior. Nevertheless, it's essential to acknowledge that such models would have inherent limitations in capturing atomic-level intricacies, making them complementary rather than a substitute for atomistic modeling in comprehensive research and understanding.

**Machine Learning Study:**

Machine Learning (ML) indeed holds great promise in addressing the computational challenges mentioned earlier. However, it comes with its own set of challenges and data requirements:

*Data Generation for ML Models:* Developing ML models for tasks like predicting interlayer formation energy and charge transfer is essential, but it necessitates the creation of a substantial dataset. Many 2D materials databases, such as C2DB[36], primarily provide information about monolayer formation energy. Generating data for interlayer properties across a wide range of 2D homo- and heterostructures is a monumental task. Machine learning analyses for interface systems are limited due to the scarcity of training data. Popular materials databases like Materials Project[35] and OQMD[37] lack information on interfaced systems, making it impossible to extract relevant data. Additionally, generating training data for interfaced systems through first-principle calculations is computationally expensive.

*Challenges in Training Data Generation:* Generating interface training data is challenging because first principle calculation of interfaced systems is computationally expensive. For example, machine learning work in Figures 24, 25 were performed with very limited data. Let's consider interfaced system in Figure 33. Let's consider DFT relaxation in 296 steps. For ML training, data at intermediate steps (50, 100, 150, 200) were considered. In this way, consider $N$ interface systems. If $n$ intermediate configurations are considered, we have total $N x n$ configurations. However, the problem of generating training data in this way is that subsequent frame may not contribute novel information. Therefore, generating training data in this way is not very reliable for accurate prediction.

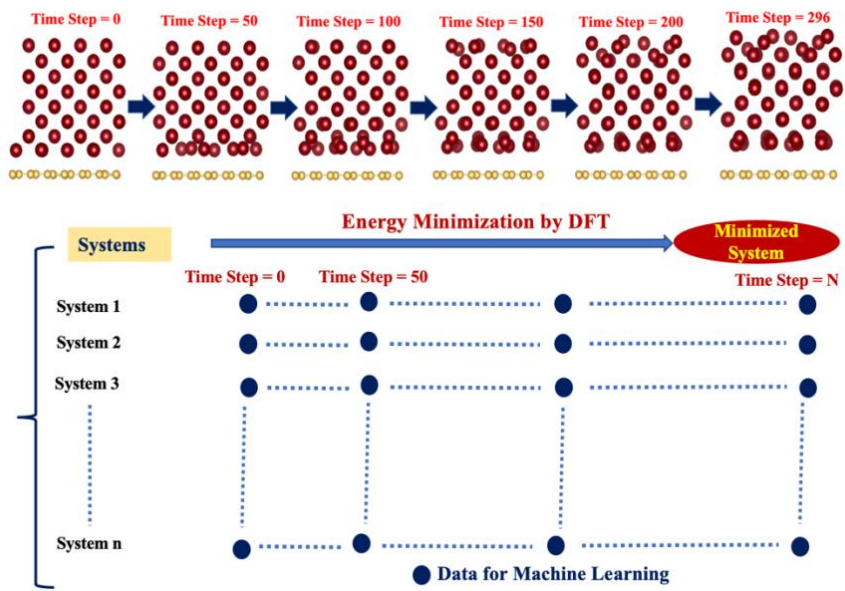

**Figure 33:** Utilizing intermediate DFT configuration files for generating Machine Learning training data.

*MD Data and Machine Learning Potentials (MLP):* For MD studies involving interlayer stability, phase transitions, friction, and fracture, suitable interatomic potentials are vital. Implementing Machine Learning Potentials[99-102] (MLPs) can expedite calculations for various intercalated systems, but developing these MLPs requires significant data generation, including DFT calculations.

Given these challenges, there is an urgent need to develop dedicated databases for interface systems. These databases should cover a range of dimensions ($n$D + $n$D, $n = 0,1,2,3$). Examples of 2D/3D interface systems like Si/graphene, Si/MoS$_2$, Si/Ti$_3$C$_2$, etc. can be included. Initially, focusing on the formation energy of interfaced systems can provide valuable information about interface adhesion strength. Furthermore, considering intercalated states, such as different stages of lithiation (Li$_x$Si) on graphene, can enhance the database's utility. To create a comprehensive database for interface systems like Materials Project,[35] it will require a concerted and large-scale collaborative effort within the research community. Such a database could greatly accelerate research in this area and facilitate the development of machine learning models for interface-related tasks.

## 6   CONCLUSIONS

The mechanical properties of 2D materials and their heterostructures have received extensive attention, especially since the discovery of graphene in 2004. However, when these 2D materials and their heterostructures are applied in electrochemical energy storage systems (EESS), their mechanical behavior is influenced by various factors, including charge/discharge rates, intercalation ions, and electrolyte types (organic and aqueous). Consequently, research on the mechanical properties of 2D materials in the context of EESS is relatively scarce or virtually non-existent.

This paper serves as a pioneering effort by shedding light on the existing challenges within EESS and how 2D materials can potentially overcome these obstacles. It delves into various aspects of studying the mechanical properties of 2D materials when applied in EESS, encompassing topics such as fracture, phase transformation, interface behavior, and interlayer interactions. Additionally, the paper underscores the computational challenges that must be overcome to tackle these open problems effectively.

One significant avenue for advancing the understanding of these systems is the development of machine learning models tailored to the study of 2D materials in EESS. To accomplish this, specific databases must be created, considering the effects of various ions, electrolyte types (both organic and aqueous), and charge/discharge rates. Building such a comprehensive database, akin to the Materials Project[35], necessitates a collaborative effort among multiple universities and the broader scientific community.

The emerging field of chemo-mechanics[103], which integrates principles from chemistry and mechanics, presents an interdisciplinary realm ripe with opportunities for research across diverse fields, including chemistry, mechanics, materials science, computational science, and more. This paper aims to catalyze computational research efforts aimed at addressing these challenges and promote collaboration with experimentalists to advance our understanding of 2D materials in EESS.

## ACKNOWLEDGEMENT

The author acknowledges the funding support from National Science Foundation (Award Number - 1911900 and 2237990) and Extreme Science and Engineering Discovery Environment (XSEDE) for the computational facilities (Award Number - DMR180013).

## COMPETING INTEREST

The authors declare no competing financial interest.